\documentclass[aps,pra,superscriptaddress,onecolumn]{revtex4-2}

\usepackage{mathtools}
\usepackage{amsmath,amscd,amssymb,amsthm}

\begin{document}

\title{Next-generation interferometry with gauge-invariant linear optical scatterers}

\author{Christopher R. Schwarze}
\email[e-mail: ]{{\tt crs2@bu.edu}}
\affiliation{Department of Electrical and Computer Engineering \& Photonics Center, Boston University, 8 Saint Mary’s St., Boston, Massachusetts 02215, USA}
\author{Anthony D. Manni}
\email[e-mail: ]{{\tt admanni@bu.edu}}
\affiliation{Department of Electrical and Computer Engineering \& Photonics Center, Boston University, 8 Saint Mary’s St., Boston, Massachusetts 02215, USA}
\author{David S. Simon}
\email[e-mail: ]{{\tt simond@bu.edu}}
\affiliation{Department of Electrical and Computer Engineering \& Photonics Center, Boston University, 8 Saint Mary’s St., Boston, Massachusetts 02215, USA}
\affiliation{Department of Physics and Astronomy, Stonehill College, 320 Washington Street, Easton, Massachusetts 02357, USA}
\author{Abdoulaye Ndao}
\email[e-mail: ]{{\tt a1ndao@ucsd.edu}}
\affiliation{Department of Electrical and Computer Engineering \& Photonics Center, Boston University, 8 Saint Mary’s St., Boston, Massachusetts 02215, USA}
\affiliation{Department of Electrical and Computer Engineering,
University of California San Diego, La Jolla, CA, USA}
\author{Alexander V. Sergienko}
\email[e-mail: ]{{\tt alexserg@bu.edu}}
\affiliation{Department of Electrical and Computer Engineering \& Photonics Center, Boston University, 8 Saint Mary’s St., Boston, Massachusetts 02215, USA}
\affiliation{Department of Physics, Boston University, 590 Commonwealth Avenue, Boston, Massachusetts 02215, USA}

\date{\today}

\begin{abstract}
  Measurement technology employing optical interference phenomena such as a fringe pattern or frequency shift has been evolving for more than a century. The systems are being designed better, and their components are being built better. But the major components themselves hardly change. Most modern interferometers rely on the same conventional set of components to separate the electromagnetic field into multiple beams, such as plate optics and beam-splitters. This naturally limits the design scope and thus the potential applicability and performance. However, recent investigations suggest that incorporating novel, higher-dimensional linear-optical splitters in interferometer design can lead to several improvements. In this work, we review the underlying theory of these novel optical scatterers and some demonstrated configurations with enhanced resolution. The basic principles of optical interference and optical phase sensing are discussed in tandem. Emphasis is placed on both familiar and unfamiliar scatterers, such as the maximally-symmetric Grover multiport, whose actions are left unchanged by certain gauge transformations. These higher-dimensional, gauge-invariant multiports embody a new class of building blocks which can tailor optical interference for metrology in unconventional ways.
\end{abstract}

\maketitle

\section{Introduction}

Optical systems have long been a popular choice for high-precision measurement and sensing. Their general use and advancements over the last century is derived from several factors. First, the intrinsic sensitivity of any wave interference is typically a small fraction of the wavelength, which for visible light is on the order of a micron. Technological achievements in lasing and detection at even the single-photon level continue to grow, all while the once room-filling tabletop optical systems are being scaled down to integrated photonic chips. 

Another core aspect of any optical system, beyond the optical sources and detectors, is the materials used to manipulate the light. Without matter, generated beams are tautologically only able to propagate through vacuum. Materials are needed to both control the characteristics of light reflecting from or transmitting through them as well as impart signatures of physical phenomena on the light for the sake of inference. Common components include beam splitters \cite{FEARN1987485, PhysRevA.40.1371}, which divide and combine beams, as well as devices to modify the orientation of the electric field \cite{Rubano:19}. Optical system building blocks like these have been optimized through decades of scientific research and engineering. Tolerances have improved, losses have decreased, and components have become more widely available and customizable, through both traditional component suppliers and chip foundries supporting nanophotonic fabrication processes \cite{6676843, Smith:23}.

While undoubtedly novel and sophisticated optical systems have since been proposed and demonstrated for metrological purposes, such as metamaterials and spatial-light modulators, these tools are often still used in conjunction with familiar components like the beam splitter and/or are used in unconventional sensing configurations. Meanwhile, the available building blocks for use in traditional interferometers have largely remained the same.

\begin{figure}[ht]
    \centering
    \includegraphics[width=\linewidth]{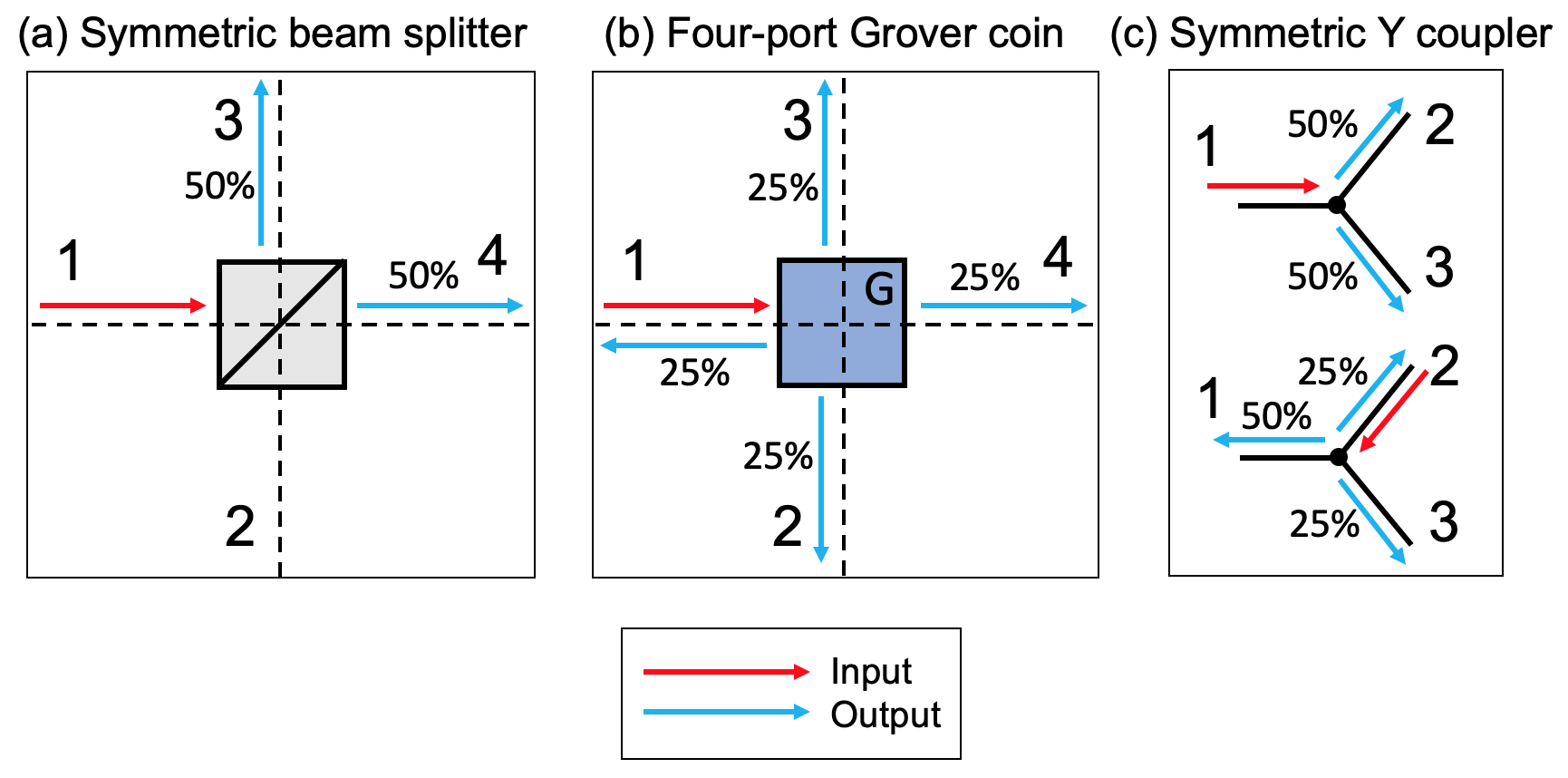}
    \caption{Comparison of some common symmetric multiports. The balanced beam-splitter (a) produces \textit{two} transmitting beams of equal, 50\% intensity for any single input, while the four-port Grover coin (b) produces \textit{four outgoing beams} of 25\% intensity, with one of those being a back-reflection. The symmetric Y coupler (c) produces two equally intense beams when input to port 1, with no back-reflection. However, when input to port 2, a 25\% back-reflection occurs, while 50\% of the energy returns to port 1 and 25\% scattering to port 3. Due to the underlying symmetry, input to port 3 is the same as input to port 2. Symmetry, in this context, is not just in the sense of having equal probabilities. The full mathematical description of the symmetries a scattering device can possess is described in Section \ref{sec:symm}. These higher-dimensional devices generalize traditional interferometric configurations; some of the examples to be discussed in detail later is shown are Fig. \ref{fig:comparison02}.}
    \label{fig:comparison01}
\end{figure}

\begin{figure}[ht]
    \centering
    \includegraphics[width=.6\linewidth]{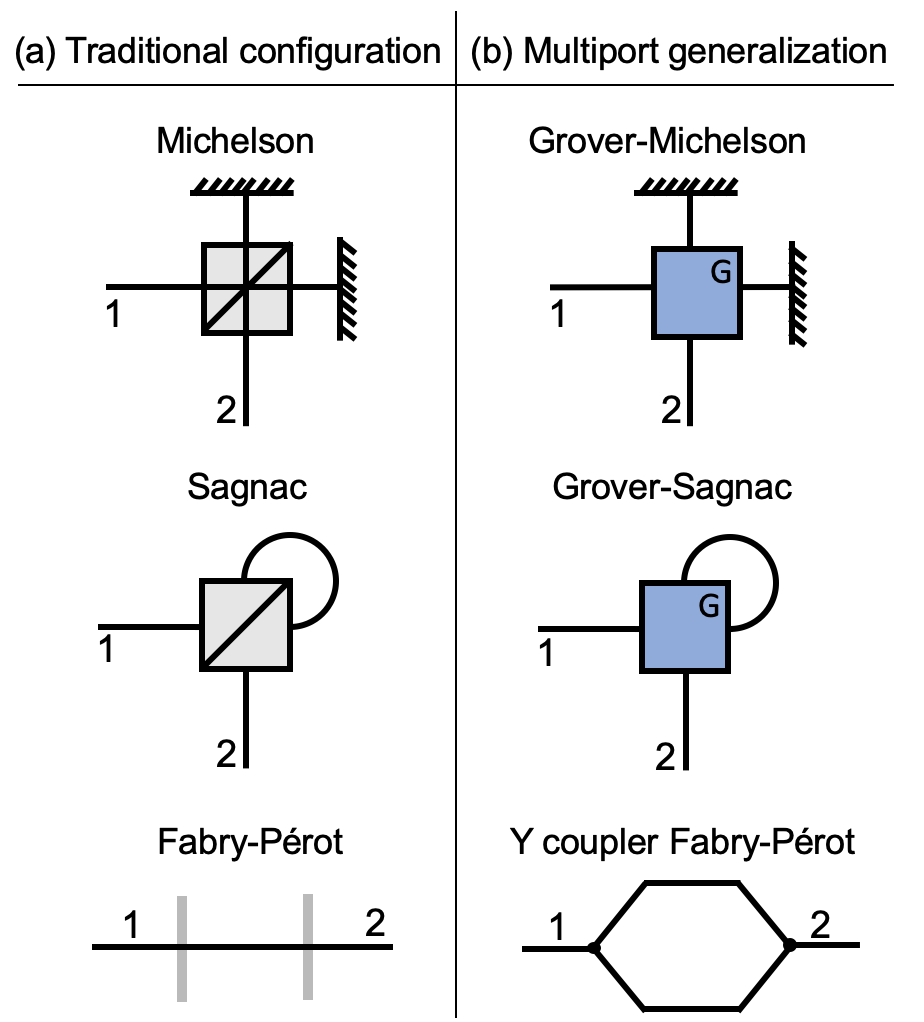}
    \caption{Examples of traditional interferometers (a) and their corresponding multiport generalizations (b). The beam-splitters in the Michelson and Sagnac configurations are replaced with the higher-dimensional Grover coin, while the two-dimensional plate optics forming the Fabry-Pérot interferometer are replaced with three-dimensional Y-couplers. In each of the multiport interferometers in case (b), the increased dimensionality combined with back-scattering behavior creates a new pair of coupled resonators. All of the pictured systems, along with other multiport interferometers, will be presented in detail in Section \ref{sec:interferometry}. }
    \label{fig:comparison02}
\end{figure}

Several alternative \textit{elementary linear scatterers} have recently been studied and demonstrated for use in optical interferometry. These devices are called higher-dimensional multiports since they actively employ more optical modes than their traditional counterparts \cite{PhysRevA.93.043845, Weihs:96, 4016, Metcalf_2013, spagnolo2013}. The most general class of these multiports admits strong coherent back-reflections, which have been demonstrated in a few cases recently \cite{Osawa:18, Kim:21, Schwarze:24}. Two important examples in this class are the four-port Grover coin\footnote{The device is named for the diffusion operator used in the quantum search algorithm of L. Grover \cite{Grover_1997}, while the term ``coin'' is to describe generic $N$-port scattering operators, especially in the study of quantum walks, since they behave like a hypothetical $N$-sided coin. The general form of an $N$-port Grover coin is specified in Section 2.} and the symmetric Y-coupler, which schematically compared to a traditional 50:50 beam splitter in Fig. \ref{fig:comparison01}. All three of these devices are interrelated and possess certain symmetries which will be discussed later in the text. The higher-dimensional, back-scattering or ``directionally unbiased'' devices have recently been shown to offer various enhancements to interferometric sensing systems, such as increased resolution, synchronous readout of several quantities, and a reduced resource count. These devices and the basic interferometric configurations they improve will form the main subject of this review. Some of the interferometric configurations to be discussed further are depicted in Fig. \ref{fig:comparison02}. In conjunction with this, we will also present the basic principles of linear optical scatterers and interferometers, emphasizing a practical mathematical framework for identifying the symmetries these devices possess.

Beyond the devices capable of dividing and recombining light, which typically entails interfacing different materials, yet another crucial building block of an optical system is ordinary transparent matter. Bulk transparent matter imparts a phase to the electromagnetic wave propagating within it. This acquired phase differs from the phase that wave would normally acquire in empty space. For a transparent region of length $d$, source wavelength $\lambda$ (in vacuum) and refractive index $n$, the phase $\phi$ acquired is
\begin{equation}\label{eq:phase}
\phi = \frac{2\pi n d}{\lambda}.
\end{equation}
This simple formula is of fundamental importance in optical phase metrology, since it illustrates precisely what quantities can alter the phase, and thus ultimately be inferred from a phase sensing instrument. The wavelength dependence makes it clear different colors naturally acquire different phases while propagating through \textit{any} medium, even vacuum. The refractive index $n$ is a dimensionless quantity that indicates the speed of light propagating through the material, or equivalently, the change in the apparent wavelength. It is entirely responsible for the difference in phase acquired inside and outside of the material. Inside a material, its value is intrinsic to its composition of matter and thus varies in a considerable number of ways. These different variations underlie much of optical metrology, enabling many physical quantities to impart an optical phase shift which can be detected.

In the next section, we present a general mathematical framework to describe how fixed distributions of optical materials scatter light. \textit{This framework abstracts away the underlying physical dynamics, treating the action on an input field energy distribution as a linear transformation enacted by the scatterer. It allows possessed symmetries of linear-optical scatterers to be identified in relation to certain field transformations, which merely represent a change in gauge and therefore do not affect any measurable quantity.} Certain symmetric scatterers such as the ones shown in Fig. \ref{fig:comparison01} emerge when their parent space is restricted to sets which remain invariant under such transformations. Thereafter in Section 3, interferometers are defined as \textit{phase-tunable} scatterers, formed from graphs. The nodes of these graphs are scatterers of the static type, and each edge connecting two scatterers is associated with an optical phase parameter. There we review the operating principles of traditional phase-sensing interferometers. This provides a baseline for contrasting the modern proposals that employ higher-dimensional multiport building blocks. After discussing these systems and their advantages, general conclusions are summarized in Section 4.

\section{Optical interference \& linear-optical scatterers} 
\subsection{Background}
The phenomenon of interference is fundamentally the result of superimposing, or adding together, different waves that exist in space. The goal of this section is to provide a mathematical description of this superposition in optical scattering devices. The optical states will be represented as a vector with complex-valued entries. The complex-number representation allows both phase and magnitude (total energy) to be encoded in each entry. Meanwhile the scatterers, which distribute the optical energy according to a linear transformation, will be represented by a matrix, which acts on these vectors. 

Generally, any wave or periodic phenomenon can be expressed or decomposed into a basis of orthogonal modes, such as the $m$-indexed functions $\sin(mx)$ and $\cos(mx)$ which underlie Fourier series expansions. In optics, the fields comprising the electromagnetic waves produced by a given source can be expanded in a modal basis which depends on the refractive index distribution considered. In a homogeneous medium, such as vacuum, these are the familiar plane wave functions of the form $E_0 e^{i\mathbf{k}\cdot\mathbf{r} - \omega t}$, where $|\mathbf{k}| = 2\pi n/\lambda$, the optical frequency $\omega$ is $2\pi c/\lambda$, $c$ is the vacuum speed of light, $n$ is the material refractive index, and $\lambda$ is the vacuum wavelength. $E_0$ is the complex-valued mode amplitude. The problem of finding the modal functions in general involves solving a complicated boundary value problem which often must be done numerically. However, it is not completely necessary to provide a useful description of optical interferometers. The present approach is one of many other prevalent matrix methods employed in optical calculations \cite{gerrard1994introduction, kim2012fourier}.

\subsection{Matrix formalism of linear-optical scattering}
We will adopt the quantum-optical notation since it generalizes to the two-photon states employed in a later section and encompasses all possible input states. We use the operator $a_j^\dagger(\omega, \boldsymbol{\varepsilon})$ to denote the creation of a photon of fixed frequency $\omega$ and polarization vector $\boldsymbol{\varepsilon}$ in spatial mode $j$. Hereafter the scope will be confined to a single frequency and polarization, so the explicit dependence of frequency and polarization of $a_j^\dagger$ can be omitted. The spatial mode index is also implicitly tied to some definite wave-vector $\mathbf{k}_j$ (plane wave model). This dependence can also be omitted without loss of generality, as long as the propagation direction is specified with an additional subscript whenever counter-propagating excitations are present within the same spatial mode.

With this notation, a state corresponding to a single photon propagating within this mode would be denoted $|\psi\rangle = a_j^\dagger |0\rangle = |j\rangle$. These single-photon basis states are orthogonal: $\langle k | j \rangle = 1$ for $k = j$ and $0$ otherwise. A general superposition of a photon across $d$ spatial modes is then denoted
\begin{equation}\label{eq:state}
|\psi\rangle = \sum_{j = 1}^d c_j|j\rangle.
\end{equation}
The coefficients $c_j$ are the photon's complex-valued probability amplitudes. The real number $|c_j|^2 = c_j^*c_j$ is then the probability of measuring the photon in mode $j$. The state is assumed normalized, so that $\sum_{j=1}^d |c_j|^2 = 1$.

By identifying the modal basis vectors $|j\rangle$ with the standard basis vector $\mathbf{e_j}$, whose $k$th element is 1 for $k = j$ and 0 otherwise, the state in the preceding equation can be placed into a state vector
\begin{equation}\label{eq:linalg}
\sum_{j = 1}^d c_j|j\rangle \longrightarrow \sum_{j = 1}^d c_j\mathbf{e_j} = 
\begin{pmatrix}
    c_1\\
    c_2\\
    \vdots\\
    c_d
\end{pmatrix}.
\end{equation}
This representation makes it clear that the total number of modes $d$ is a measure of dimension. Optical qubit states employ two modes. States involving coherent superpositions of $d$ modes are sometimes known as qudits.

The classical equivalent of a state in the form of Eq. (\ref{eq:state}) is found by replacing the creation operator with the classical field mode function. This rescales all of the amplitudes by the square root of the total power in the beam, converting the probabilities into optical power or a proportionally related quantity. Hence, the probability distribution $|c_j|^2$ vs $j$ is effectively a normalized distribution of optical energy in the state $|\psi\rangle$. Besides that rescaling, in terms of the algebraic formalism, it only shifts what is being tied to the standard basis $\{\mathbf{e_1}, \dots, \mathbf{e_d}\}$. This has no impact on the description of the scattering devices and thus the resulting interference they induce. Any distinction made between single-photon states and classical coherent states in the context of linear interferometry is largely inconsequential other than an overall normalization factor. Ref. \cite{Barnett_2022} contains further commentary on this. 

\begin{figure}[ht]
    \centering
    \includegraphics[width=0.5\linewidth]{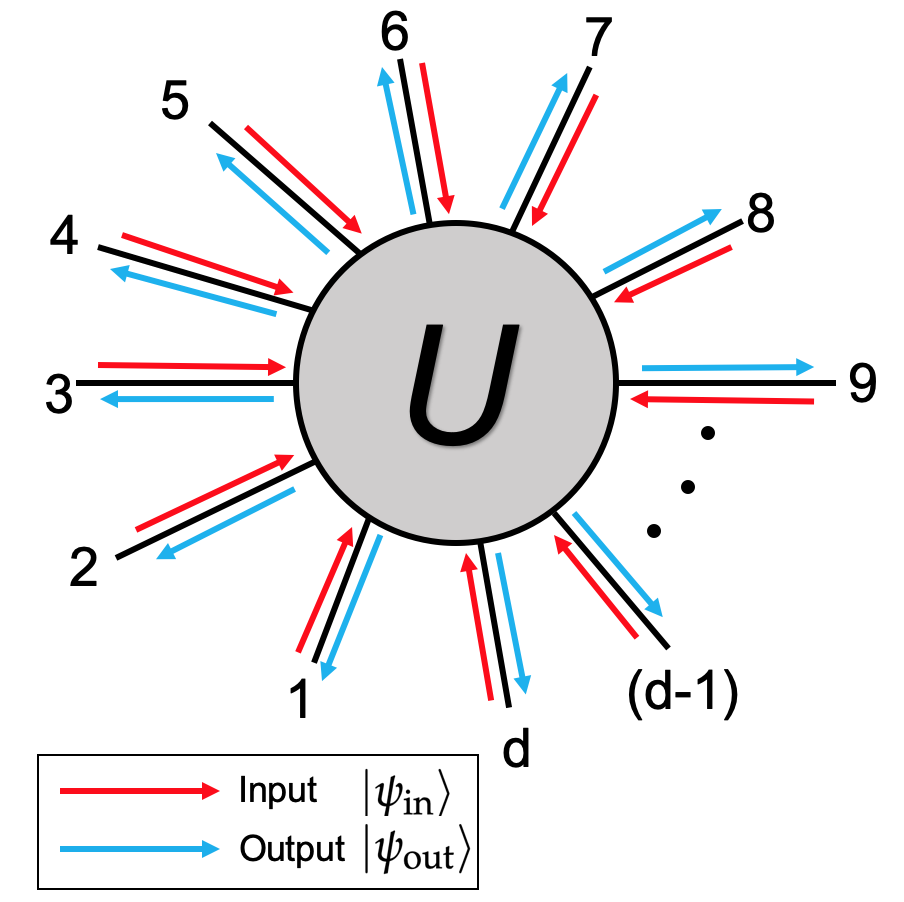}
    \caption{A generic, lossless linear-optical scattering device. This $d$-port device is mathematically represented by a $d\times d$ unitary scattering matrix $U$. Port $j$ is identified with two spatial modes, one for each direction of propagation. The scatterer converts an input optical state $|\psi_{\text{in}} \rangle$ to the output state $|\psi_{\text{out}} \rangle = U| \psi_{\text{in}} \rangle$. This scattering action redistributes the optical energy in the input state. The arrows represent the flow of the associated state amplitudes. For clarity the counter-propagating arrows at the same port are drawn spatially separated, but in reality, the counter-propagating energy flow is perfectly collinear in each spatial mode. Scattering devices like $U$ can be viewed as nodes of a general graph. Depictions like the above and more traditional nodal depictions such as those in Fig. \ref{fig:equiv} will be used interchangeably. A graph of scattering devices is generally called an \textit{interferometer}.}
    \label{fig:scatter}
\end{figure}

A linear scatterer will act on the incident state $|\psi_{\text{in}}\rangle$ according to a $d\times d$ matrix transformation $U$, producing the state $|\psi_{\text{out}}\rangle = U|\psi_{\text{in}}\rangle$. This implicitly reverses the direction of all the modes, as depicted by the general scattering situation in Fig. \ref{fig:scatter}. $U$ is known as the scattering matrix. When a photon located entirely in mode $j$ enters the corresponding port of $U$, the state produced, in the form of Eq. (\ref{eq:linalg}), is simply column $j$ of $U$. This implies that the element $U_{ij}$ is the probability amplitude for a photon entering port $j$ of the scattering device to emerge at port $i$. Thus, $|U_{jj}|^2$ is the probability for a photon entering port $j$ to reflect directly backwards. The corresponding amplitudes for this back-scattering lie on the diagonal of $U$.

When a photon impinges port $j$, the requirement of energy or probability conservation entails that $\langle \psi_{\text{out}} | \psi_{\text{out}} \rangle = \sum_{i=1}^d |U_{ij}|^2 = 1$, which must then hold for all $j$. Maintaining orthogonality at the output, across different single-photon inputs, similarly requires that $\sum_{j=1}^d U_{ij}U_{jk}^* = 0$ whenever $i \neq k$. These restrictions, when combined, imply that $U^\dagger U = I$, where $U^\dagger$ is the Hermitian conjugate or complex-transpose of $U$, $U^\dagger = (U^T)^*$. This condition is equivalent to the columns of $U$ forming an orthonormal basis of $\mathbb{C}^n$. In other words, energy-conserving linear-optical scatterers are represented by unitary matrices. This subset of matrices forms a group, known as the unitary group in $d$-dimensions and will be denoted $U(d)$.

For a general input superposition, the scattering matrix $U$ redistributes the inputted energy into a new outgoing superposition. This redistribution of energy is the harbinger of general interference, allowing energy from separate beams to be combined into one. The net outcome of this redistribution will be a function of both the scattering matrix $U$ and the initial state amplitudes, and in particular, the relative phases of those state amplitudes. 

This discussion pertains to an arbitrary but fixed frequency and polarization, making $U$ a constant. Nonetheless, the scattering behavior of photons of different frequencies and/or polarizations will generally vary due to the variation of a material's optical response with respect to changes in these quantities. This in turn causes $U = U(\omega, \boldsymbol{\varepsilon})$ to be a matrix-valued function. In fact, since the material scattering response will generally be affected by any change in the refractive index distribution of the material comprising the scatterer, $U$ will as well. Optical scattering devices are often engineered to resist this variation.

Enforcing symmetries by requiring that $U$ remains invariant under a certain field transformation can impose constraints on $U$ \cite{PhysRevA.105.023509}. An example is \textit{reciprocity}, which originates in name from the Lorentz reciprocity theorem. This result pertains to a transformation that interchanges the sources and fields in an enclosed region. A reciprocal scatterer is defined as being invariant to interchange of source and detector. Recalling that $U_{ij}$ is the probability of detecting a single photon at port $i$ when it enters port $j$, we see that reciprocity stipulates that $U_{ij} = U_{ji}$, or that $U$ is a symmetric matrix: $U = U^T$ \cite{DEAK20121050}. 

\textit{Non-reciprocal} scattering devices are far less common than reciprocal ones. They often are formed by locally breaking the time-reversal symmetry, such as with a Faraday rotator. A global rotation of the system can also produce a non-reciprocal phase accumulation, via the Sagnac effect \cite{RevModPhys.39.475}. In these devices, forward-propagating light will acquire a phase which differs from backward-propagating light in the same mode. Transmission can also be non-reciprocal. For instance, a circulator provides non-reciprocal routing of light. An ideal three-port circulator can be prescribed a scattering matrix
\begin{equation}\label{eq:circ}
U = 
\begin{pmatrix}
    0 & 0 & 1\\
    1 & 0 & 0\\
    0 & 1 & 0 
\end{pmatrix}
\neq U^T.
\end{equation}
A physical device with this scattering matrix would be able to perfectly convert counter-propagating excitations in one spatial mode into identical excitations that populate separate spatial modes, and vice versa. However, circulators in practice tend to experience both losses and crosstalk. Circulators are readily defined in higher dimensions, and can be obtained by chaining together low dimensional ones.

\subsection{Scattering process gauge symmetries}\label{sec:symm}
Maxwell's equations are known to possess a certain property known as gauge invariance. This property allows the underlying field potentials to be redefined in a certain way without changing the physical situation. The principle has since been extensively generalized and utilized in other areas of physics. Generally speaking, the transformations between different choices of gauge form a symmetry group. In the following subsections, we describe analogous transformations to the optical states and/or scattering devices. These are reminiscent of the gauge transforms employed in general field theories mainly in the sense that they do not physically alter the underlying process's description. Any given gauge can be used freely for the sake of convenience. The particular forms of the transformation are different, so while they might be referred to more precisely as a scattering center symmetry or \textit{generalized} gauge transformation, we will use the generic terminology ``gauge transformation'' as a shorthand. The first set of scattering gauge transformations considered also forms a group of its own, \textit{the symmetric group} $\text{S}_d$.

For radiation of fixed frequency and polarization expanded into a basis of orthogonal spatial modes, as in Eq. (\ref{eq:state}), there are two basic transformations which do not change the physical situation \cite{PhysRevA.52.4853}. These each provide a formal equivalence relation which can be used to relate scattering matrices. Scattering devices within the same equivalence class will not be measurably distinguishable, but can appear as completely different scattering matrices. The different scattering matrices can be viewed of as the same underlying physical action but expressed in a different gauge. 

First, the physical description of any optical scattering process would be unchanged if the numerical labels used to identify the field modes were modified, for these labels are completely arbitrary. Formally, the relabeling process enacts a permutation $p$ on the creation operator indices: $a_j^\dagger \rightarrow a_{p(j)}^\dagger$. This is equivalent to relabeling the ports of a given scattering device like that shown with its labels in Fig. \ref{fig:scatter}. This variant of transformation will be discussed in Section \ref{sec:perm}.

The second transformation is the application of a phase shift to each field mode. In a general state of the form of Eq. (\ref{eq:state}), this will enact the change $c_j \rightarrow e^{i\phi_j} c_j$. Since $(e^{i\phi_j} c_j)^*(e^{i\phi_j} c_j) = |c_j|^2$, this transformation does not affect the distribution of detection probabilities, which is the physical observable. This transformation can be applied locally to any given scattering device in a general network. Further discussion of this class of transformations is deferred to Section \ref{sec:ext}. After presenting an overview of the two interrelated transformations and their general implications, the subsequent sections discuss specific consequences and examples for the scattering groups $U(2)$, $U(3)$, and $U(4)$. 

\subsubsection{Geometric scattering transformations}\label{sec:perm}
A permutation or relabeling of $d$ objects can be defined by a bijective function $p$ which maps the set of original labels $\{1, \dots, d\}$ onto itself. There are generally $d!$ such maps, and each can be represented by a $d \times d$ matrix $P$ which acts on a column vector of labels. Throughout this work we will always denote the permutation function and its corresponding matrix by an upper and lowercase pair of the same symbol. Each permutation $P$ has a well-defined inverse given by $P^T$, which ``undoes'' the permutation enacted by $P$. A full treatment of permutations can be found in any standard algebra text, such as Ref. \cite{pinter}.

For a general $d$-mode optical scattering state, permutation of the mode labels acts in the following way
\begin{equation}
|\psi\rangle = \sum_{j = 1}^d c_j|j\rangle \rightarrow |\psi'\rangle = \sum_{j = 1}^d c_j|p(j)\rangle
\end{equation}
The energy distribution encoded by this state is unchanged. It is merely being rewritten in a new basis provided by $P$. Conversely, had we simultaneously permuted the $c_j$ to $c_{h(j)}$ for a permutation $h$, which may or may not equal $p$, then the spatial energy distribution would have been measurably different before and after the permutation. This is why only the labels and not the coefficients are permuted.

Element $U_{ij}$ of a general scattering matrix is the probability amplitude for an incident photon at port $j$ to scatter to port $i$. Permuting the mode labels affixed to each port maps $U_{ij} \rightarrow U_{p(i)p(j)}$. This permutes both the rows and columns by the same permutation $p$. 

It can be shown that pre-multiplication of $U$ by the associated permutation matrix $P$ will permute the rows of $U$. Then, post-multiplying $U$ by $P^T$ will enact the permutation $p$ on the columns of $U$, which can be deduced by expressing $UP^T = (P U^T)^T$. Combining the two results, we see that the aggregate affect of a mode or port relabeling on a scattering matrix $U$ is the similarity transformation 
\begin{equation}\label{eq:symm}
    U \rightarrow PUP^T. 
\end{equation}
This is the general form of the geometric gauge transformation; that is, it describes how to transform the matrix $U$ written in one gauge into a different gauge. Being a conjugation transformation, it allows any given $U$ to be associated with the equivalence class $\{ PUP^T : P\text{ is a permutation matrix} \}$. Any members common to the same equivalence class will not be physically distinguishable, despite the general possibility that the form of their scattering matrices will differ. In accord with this, the equivalence of two scattering matrices $U_1$ and $U_2$ will be notated $U_1 \sim U_2$. 

\begin{figure}[ht]
    \centering
    \includegraphics[width=\linewidth]{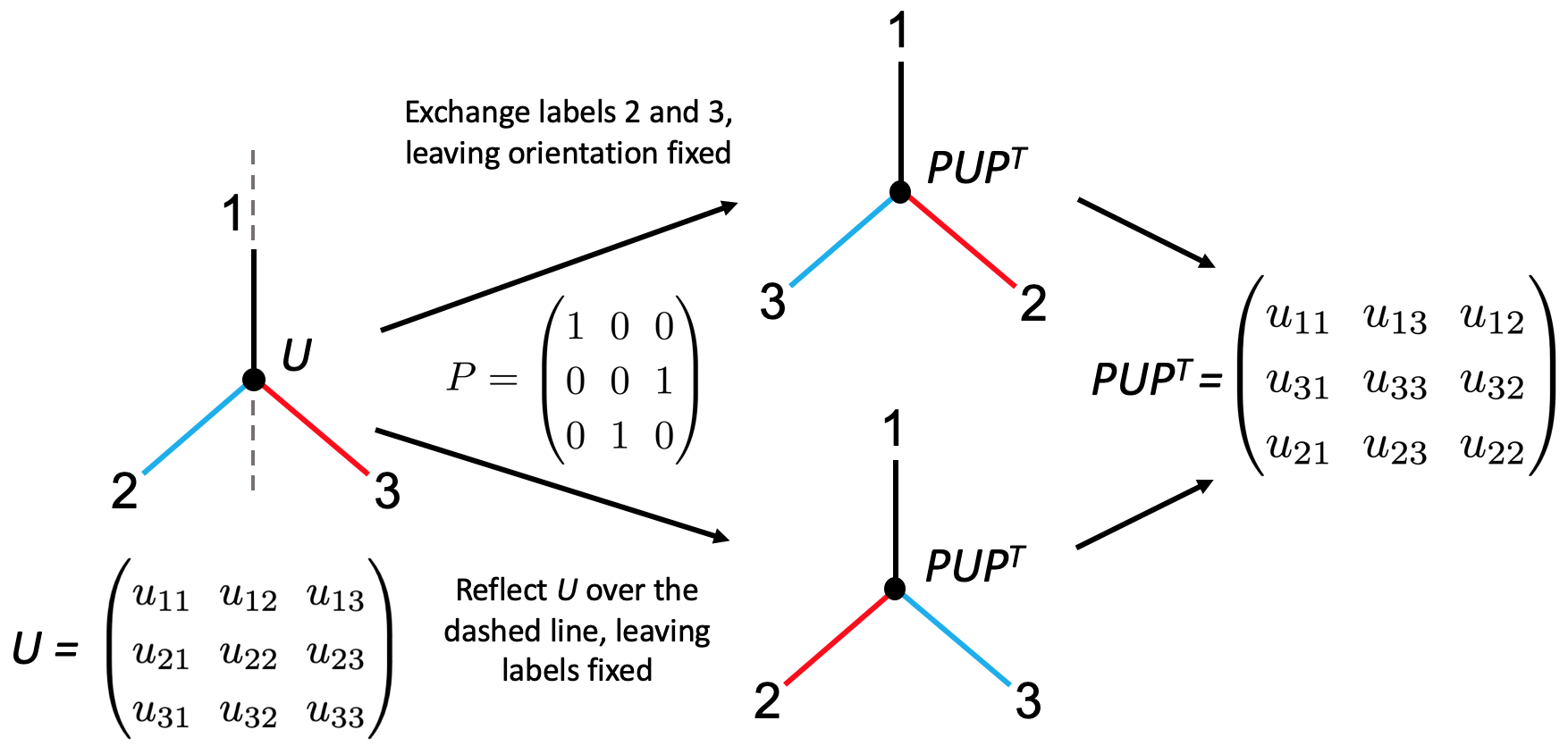}
    \caption{Geometric gauge transformations involve relabeling the field modes identified with the ports of a scattering device $U$. The indices $j$ of the mode operators undergo a permutation transformation $p(j)$, which transforms the scattering matrix $U$ to $PUP^T$, where $P$ is the permutation matrix associated with the permutation function $p$. Permuting the port labels is also equivalent to applying geometric transformations to the scattering device while leaving the labels themselves fixed. Any example is shown here for a generic three-port scattering device, where exchange of two labels equates to reflection about the line between them. The edge colors at the red and blue ports are used to signify the orientation of this pointlike scatterer $U$. General permutations $P$ in higher dimensions will enact reflections and/or rotations in higher-dimensional spaces.}
    \label{fig:equiv}
\end{figure}

The gauge freedom provided by this transformation is the ability to label the ports or modes in a way which produces the form most convenient to work with. In tandem it provides a means to show \textit{two apparently different scatterers are in fact the same}. Beyond this, though, it naturally leads to the notion of a symmetric scatterer. These so-called gauge invariant devices are left unchanged by a specific permutation. This implies $U = PUP^T$. After post-multiplying by $P$ and rearranging, this is equivalent to
\begin{equation}\label{eq:gi}
    UP - PU \coloneqq [U, P] = 0.
\end{equation}
Scattering matrix families can be defined by this condition with certain $P$ in mind, greatly reducing the number of free parameters they possess. Specific examples of this will be provided later in this section. In essence, the gauge-invariance corresponds to certain subsets of ports behaving identically when excited by the same input state. This can be connected to physical symmetries of the scattering process, such as mirror-reflection symmetry and/or rotational symmetry. This is because that \textit{permutation operations, such as exchanging or cyclically permuting subsets of the mode labels, are equivalent to reflecting or rotating the physical device while leaving the original labels fixed.} An example illustrating this equivalence notion with a generic three-port device is illustrated in Fig. \ref{fig:equiv}. 
Like any other scattering matrix, a gauge-invariant device can only be physically specified up to its equivalence class. Applying different gauges to a specific gauge-invariant device will not change the fact that it commutes with some permutation matrix, but the commuting matrix itself can change. It is straightforward to show that for $U_0$ which is gauge-invariant with respect to $P_0$, after expressing $U_0$ as $U$ in some new gauge $Q$, then the corresponding invariant-inducing permutation is transformed to $P_0 \rightarrow Q P_0 Q^T = P$. This leads to the concept of equivalent permutation classes, which are the conjugacy classes $\{QPQ^T : Q, P \text{ are permutation matrices}\}$. Members of the same class generate the same \textit{kind} of gauge-invariant scattering device. An intuitive picture of these permutation classes can be built with the equivalence picture of Fig. \ref{fig:equiv} in mind. First, a gauge-invariant scatterer is viewed as a device that is left unchanged by a geometrical transformation acting on it while its port labels stay fixed. This geometric transformation is tied to an underlying permutation $P$. Then the equivalent permutations of $P$ are merely re-expressions of the original port labels which are used to label $U$ before and after the geometrical transformation. Clearly, this relabeling has no effect on the geometrical transformation encoded by $P$. 

The transformation enacted by Eq. (\ref{eq:symm}) possesses several basic algebraic properties, which all reflect the fact that it cannot change the physical description of the scattering process $U$ represents. The gauge transformation will always preserve the reciprocity and unitarity of $U$. Another important example concerns the spectrum of $U$ as a gauge transformation is made. In that case, the eigenvalues are unchanged while the eigenvectors corresponding to these eigenvalues are permuted by $P$.

Permutation invariance can also be applied to optical states. Since the phases of the state amplitudes are immaterial when the state undergoes detection, these can be discarded. Then, a symmetric scattering state is defined by having an energy distribution which is left unchanged by some gauge transformation $P$. This is the type of symmetry usually associated with balanced scatterers that produce beams of equal intensity when a single beam impinges them. Gauge invariance of the scattering device alone is generally unrelated to the creation of symmetric states.

\subsubsection{External phase shifts}\label{sec:ext}

In addition to relabeling the field mode indices, which are directly tied to a scattering device's port labels, external phase shifts placed around a scattering device will have no affect on its measured energy distribution when a photon enters one of its ports. External phases translate the total phase acquired by the photon in each mode $j$ by some constant $\phi_j$. However, the input excitations have no absolute reference phase, so this translation will not be measurable, due to a lack of knowledge of the starting point.\footnote{A non-reciprocal phase is a relative entity, shared between the counter-propagating directions of the same spatial mode. Therefore an external non-reciprocal phase is be able to have a measurable effect on scattering systems in general. For example, consider a standard Mach-Zehnder interferometer, where a tunable phase shift $\phi$ in one arm is used to produce an interferogram. An external phase shift placed in that arm will naturally add with $\phi$. If it is reciprocal, this just translates $\phi$ and the interferogram it produces, which has no absolute starting point anyways. But, an external non-reciprocal phase will generate measurable non-reciprocal transmission in a device comprised of entirely reciprocal scattering devices.} 

The general action of a set of externally placed phase shifts is mapping the input state coefficients $c_j \rightarrow c_j e^{i\phi_j}$, which can be encoded as a pre-multiplication by the diagonal matrix $D = \text{diag}(e^{i\phi_1}, \dots, e^{i\phi_d})$. This means when scattering through the device with the external phases, $|\psi_{\text{in}}\rangle \rightarrow DUD |\psi_{\text{in}}\rangle$, or equivalently, 
\begin{equation}
    U \rightarrow DUD.
\end{equation}
This induces another equivalence notion between different scattering matrices. Two scattering matrices $U_1$ and $U_2$ are equivalent in this sense whenever a suitable selection of external reciprocal phases converts one scattering matrix into the other. In other words, this gauge freedom allows arbitrary selection of desired reciprocal external phases to recast $U$ into the different but equivalent form $DUD$. It is readily verified that changing the gauge in this form will neither affect unitarity nor reciprocity. 

All elements $U$ in the equivalence class $\{ DUD : \text{diag}(e^{i\phi_1}, \dots, e^{i\phi_d}), \phi_1, \dots, \phi_d \in \mathbb{R}\}$ necessarily have the same probability matrix, formed by taking the elementwise modulus square of $U$. In some special cases, probabilities will be zero, allowing the corresponding phases to be chosen freely. This can enable some more non-trivial equivalences. On the other hand, the notion of gauge invariance has little use here, since it would require $DUD = U$. This is generally only valid when the external phases are brought to zero, so that $D = I$. This case is omitted, as it was for $P = I$ in the previous section, since all $U$ are equivalent to a matrix which is gauge-invariant in this sense. For this reason, the term gauge invariant is only used in the context of nontrival (i.e. $P \neq I$) geometric gauge transformations throughout this work.

Although this phase gauge freedom can be invoked in conjunction with the geometric gauge freedom, the effects of these two gauge freedoms are not completely separate. Scattering devices which are gauge invariant in the geometric sense will lie in the same phase-gauge equivalence classes as devices which are not, since these phases are easily reselected to break any given geometric symmetry. External phases also will generally change the eigenvectors of the scattering device they surround. Notwithstanding, considering the two gauge freedoms together allow fairly nontrivial equivalences to be made. Establishing such an equivalence in turn implies that interferometers formed by substituting one scatterer for its equivalent counterpart will produce the same interferometer, up to global translations of its parameter space.

Besides writing the external phase gauge transformation as $U\rightarrow DUD$ for a diagonal phase matrix $D$, the map $U_{ij} \rightarrow U_{ij}e^{i\phi_i + i\phi_j}$ admits another useful form. Let $e^{i\phi_i + i\phi_j} = e^{i\phi_i}e^{i\phi_j} = v_i v_j$ be written as the outer product $vv^T$. When external phases are selected for a scattering matrix $U$, $U$ is element-wise multiplied by the corresponding matrix $vv^T$, denoted $U \odot vv^T$.

This form, $U \rightarrow U \odot vv^T$ for $v_j = e^{i\phi_j}$ can be more practical than the first one; it provides an algorithm to quickly determine whether two scatterers are equivalent or not, and if they are, what phases are required to construct the explicit gauge transformation between them. Let $U_1$ and $U_2$ be scattering matrices with the same classical probability matrix. Express $U_1 = |U| \odot Q_1$ and $U_2 = |U| \odot Q_2$ where $|U|$ is the matrix formed by taking the magnitude of each element of $U_1$ and $U_2$. Next define $Q = Q_1 \odot Q_2^{\circ -1}$, where $Q_2^{\circ -1}$ is found by taking the scalar inverse of each element of $Q_2$. The matrices $Q_1$ and $Q_2$ will be entirely phase factors, so this inverse is always well-defined. Since $U_1$ and $U_2$ have the same underlying probability matrix, $Q = U_1 \odot U_2^{\circ -1}$. By the above, if $Q$ is expressible as an outer product, then $U_1 \sim U_2$, and $Q$ directly prescribes the gauge transformation as $U_2 \odot Q = U_1$. 

For an outer product matrix, each column must be a scalar multiple of the first column. This is easy to check, can be done in parallel, and can be halted upon any column's failure to meet the condition. If $Q$ meets the condition, the outer product factorization $vw^T = Q$ is found during the assessment stage, since $v$ will form the first column of $Q$ while the very factors that scale this column to generate each other column will form $w$. A global scale factor $1 = e^{i\theta-i\theta}$ can then be used to exchange phase between $v$ and $w$ to symmetrize the outer product into the form $vv^T$. 

\subsection{Directionally biased and unbiased devices}
Now that the background theory for the two scattering gauge freedoms has been established, specific systems can be considered in detail. When an input scattering state $|\psi_{\text{in}}\rangle$ is directed toward a gauge-transformed $U' =  PUP^T$, the output state is $|\psi_{\text{out}}\rangle = PUP^T |\psi_{\text{in}}\rangle$. This product can be intuitively described in the following steps: first the input state is first transformed by $P^T = P^{-1}$ to undo the permutation, then it experiences the scattering action of the original $U$, and then it is placed back into the correct labeling with $P$. 

When these permutations are applied to the indices of the scattering state's mode operators, it has been implicitly assumed that the input modes and output modes were the same. This is a direction-free or \textit{directionally unbiased} convention. Strictly speaking the incident state basis is different than that of the scattered state leaving the device, since the modes corresponding to counter-propagating optical excitations in a given spatial mode are distinct. It is often unambiguous to disregard this distinction, but in instances where counter-propagating excitations populate the same spatial mode, the direction needs to be specified. For instance $a_{jF}^\dagger$ can represent forward propagation while $a_{jB}^\dagger$ represents backward propagation. This would effectively double the dimension of all the matrices involved: the operator basis would go from $\{a_1^\dagger, \dots, a_d^\dagger\}$ to $\{a_{1F}^\dagger, \dots, a_{dF}^\dagger, a_{1B}^\dagger, \dots, a_{dB}^\dagger \}$ where $F$ represents ingoing and $B$ represents outgoing. 

By definition, the only scattering that occurs is cross-coupling between the ingoing and outgoing modes. This means any scattering or permutation matrix $A$ expressed in the direction-free basis would become a block matrix
\begin{equation}\label{eq:dirspecific}
    A\rightarrow
    \begin{pmatrix}
        0 & A^*\\
        A & 0\\
    \end{pmatrix}
\end{equation}
in the direction-specific basis. The conjugate is needed to express the coupling going backwards, or from outgoing to ingoing, in order to maintain time-reversal symmetry. The dimensionality is not changed in either representation: the block $2d$ system has dimensionality $d$, so for the sake of convenience the compact $d \times d$ notation is used. 

\begin{figure}[ht]
    \centering
    \includegraphics[width=\linewidth]{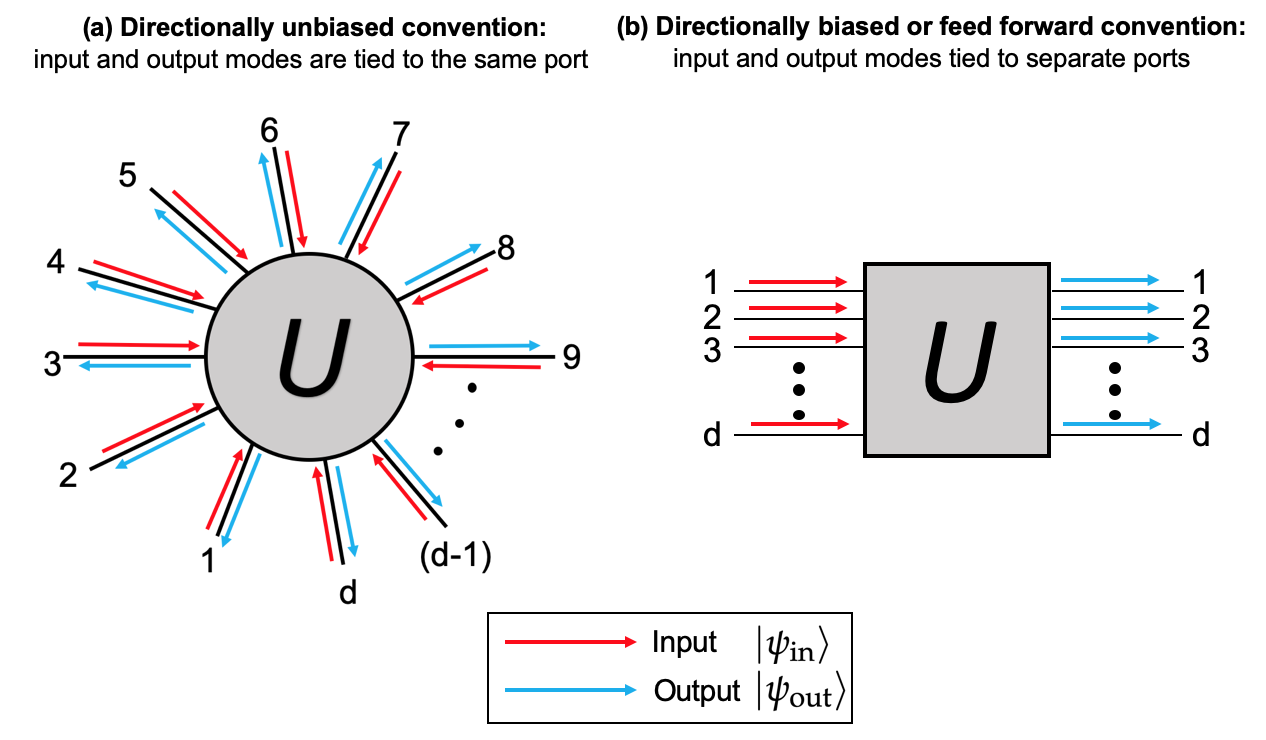}
    \caption{Two equivalent scattering matrix conventions. In convention (a) the ingoing and outgoing modes are associated with the same port, producing a $d\times d$ matrix $U$. In convention (b) these sets of modes are kept distinct, and assigned to different ports. This produces the $2d \times 2d$ block matrix form in Eq. (\ref{eq:dirspecific}). The ports are abstract entities, so their identification with a single mode or pair of counter-propagating modes is immaterial. Nevertheless, certain, so-called block feed-forward scattering matrices expressed in the unbiased convention mimic the block form, and can be compactified further, as in Eq. (\ref{eq:compact}).}
    \label{fig:dub}
\end{figure}

The sparse notation in the preceding equation can be viewed as a feed-forward or feed-through system, where the scattering device routes all of the energy in one set of modes through to another disjoint set of modes. An analogous reduction in dimensionality can occur once again, in the direction-free picture, under the assumption of the following gauge symmetry. Let $d$ be an even number and $U$ be a $d \times d$ scattering device (expressed in the compact notation). Assume $U$ commutes with the gauge transform operator $P$ where 
\begin{equation}
    P = 
    \begin{pmatrix}
        0 & I\\
        I & 0
    \end{pmatrix}
\end{equation}
and $I$ is the $d/2$ identity matrix. Expressing $U$ in the block form
\begin{equation}\label{eq:Ugen}
    U = 
    \begin{pmatrix}
        A & B\\
        C & D
    \end{pmatrix}
\end{equation}
and enforcing $PUP^T = U$ leads to the conditions $A = D$ and $B = C$, so that $U$ can be written 
\begin{equation}
U = 
\begin{pmatrix}
    A & B\\
    B & A
\end{pmatrix}.
\end{equation}
This is the general form of a scattering device endowed with the gauge symmetry $P$. A \textit{block feed-forward} scattering device additionally satisfies the condition $A = 0$. This produces the matrix 
\begin{equation}\label{eq:compact}
U = 
\begin{pmatrix}
    0 & B\\
    B & 0
\end{pmatrix}.
\end{equation}
Thus, a block feed-forward matrix scatters all light in the set of modes $\{a_1^\dagger, \dots, a_{d/2}^\dagger\}$ into the disjoint set$ \{a_{d/2 + 1}^\dagger, \dots, a_d^\dagger\}$, mimicking the form of Eq. (\ref{eq:dirspecific}). Sometimes these scattering devices are called \textit{directionally biased}, since they admit a unidirectional sense of energy flow, which is illustrated in Fig. \ref{fig:dub}. These block feed-forward devices are a special case of the general non-reflecting or feed-forward scattering device, which is defined by the condition that its diagonal values are all zero. The resemblance to Eq. (\ref{eq:dirspecific}) is even closer when the transformation is reciprocal, as this allows $U$ to be written
\begin{equation}\label{eq:compactreciprocal}
U = 
\begin{pmatrix}
    0 & B^T\\
    B & 0
\end{pmatrix}.
\end{equation}
Reciprocal block feed-forward devices are ubiquitous. They allow the $d\times d$ matrix $U$ to be represented by an even further compactified \textit{transmission matrix} $B$. $B$ inherits the unitarity of $U$, so block feed-forward scatterers can be viewed as the embedding of a lower-dimensional unitary group in a higher-dimensional space. This change in the effective dimensionality is tied to the underlying gauge symmetry, which itself results from the system possessing a mirror symmetry with respect to the entire sets of input and output modes. 

Even if $U$ is reciprocal, the block $B$ itself does not need to satisfy $B = B^T$. This allows block feed-forward devices to produce effectively non-reciprocal mathematical operations. The aforementioned gauge freedoms also may be applied to $B$. In fact, the condition $B = B^T$ can be tied to another gauge invariance, corresponding to block mirror symmetry between the input and output sets of modes.

Block feed-forward scattering devices include the standard beam-splitter, the waveguide coupler, as well as the Reck and Clements beam-splitter meshes \cite{PhysRevLett.73.58, Clements:16}. Multi-mode waveguide interferometers and $d\times d$ waveguide couplers such as a fiber tritter also realize this form \cite{Suzuki:06}. The Hadamard-convention beam-splitter transmission matrix $H$ is 
\begin{equation}\label{eq:H1}
    H = \frac{1}{\sqrt{2}}
    \begin{pmatrix}
        1 & 1\\
        1 & -1
    \end{pmatrix}
\end{equation}
whereas its regular scattering matrix is 
\begin{equation}\label{eq:H2}
    \begin{pmatrix}
        0 & H\\
        H & 0
    \end{pmatrix}.
\end{equation}
The three-port optical circulator introduced in Eq. (\ref{eq:circ}) serves as the physical link between the feed-forward and unbiased forms of any given scattering device. Placing a circulator at each port of an unbiased device will produce the equivalent block feed-forward one. To convert a block feed-forward device into an unbiased one, each input/output pair needs to be connected to a common three-port circulator.

\subsection{\textit{U}(1) scattering processes}

A $U(1)$ scattering process is simply a back-reflection that acquires some necessarily reciprocal constant phase $e^{i\phi}$. This phase can always be gauged away using the external phase shift freedom. The only permutation available is the trivial identity permutation, so no interesting geometric gauge invariants exist in this space. Overall this scattering space is trivial, containing just the identity equivalence class.

\subsection{\textit{U}(2) scattering processes}\label{sec:U2}

For a generic $U(2)$ scattering device $U$ with external phase shifts $\phi_j$ at port $j$, light entering either port can either transmit through or reflect back. The general form of this transformation (for $\phi_1 = \phi_2 = 0)$ is thus
\begin{equation}\label{eq:U2}
U = 
\begin{pmatrix}
    r_1e^{i\delta_1} & t_2e^{i\gamma_2}\\
    t_1e^{i\gamma_1} & r_2e^{i\delta_2}
\end{pmatrix}.
\end{equation}
The numbers $r_1, r_2, t_1, t_2$ are the magnitudes which lie in the range $[0, 1]$, while $\delta_1, \delta_2, \gamma_1, $ and $\gamma_2$ are the corresponding real-valued phases. Due to unitarity, we have normalization conditions $r_1^2 + t_1^2 = r_2^2 + t_2^2 = 1$ as well as orthogonality, manifesting as
\begin{equation}
r_1t_1 e^{-i\delta_1 + i\gamma_1} + r_2t_2 e^{i\delta_2 - i\gamma_2} = 0.
\end{equation}
This can be rewritten as
\begin{equation}
r_1(1-r_1^2)^{1/2} + r_2(1-r_2^2)^{1/2} e^{i\phi} = 0, 
\end{equation}
where $\phi = \delta_1 + \delta_2 + \gamma_1 - \gamma_2$. Both $r_1(1-r_1^2)^{1/2}$ and $r_2(1-r_2^2)^{1/2}$ are positive real numbers, implying that the variable $\phi$ must be selected such that $e^{i\phi} = -1$, yielding 
\begin{equation}
r_1(1-r_1^2)^{1/2} = r_2(1-r_2^2)^{1/2}.
\end{equation}
This can only be satisfied for $r_1 = r_2$ (meaning $t_1 = t_2)$ as well. Thus, a two-dimensional unitary scattering device \textit{cannot have non-reciprocal transmission}. The phase constraint is that 
\begin{equation}\label{eq:HOM}
\phi = \delta_1 + \delta_2 - \gamma_1 - \gamma_2 = 2\pi m + \pi
\end{equation}
for any integer $m$. This constraint is unchanged when external phases are included, but including those allows us to set $\phi_1 = -(\gamma_1 + \gamma_2 - \delta_2)/2$ and $\phi_2 = -\delta_2/2$, so that $\phi_1 + \phi_2 = -(\gamma_1 + \gamma_2 - \delta_2)/2 -\delta_2/2 = -(\gamma_1 + \gamma_2)/2$, resulting in
\begin{equation}\label{eq:reduced}
\begin{pmatrix}
    -r & (1-r^2)^{1/2}e^{i(\gamma_2 - \gamma_1)/2}\\
    (1-r^2)^{1/2}e^{-i(\gamma_2 - \gamma_1)/2} & r 
\end{pmatrix}
=
\begin{pmatrix}
    -r & (1-r^2)^{1/2}e^{i\Delta}\\
    (1-r^2)^{1/2}e^{-i \Delta} & r 
\end{pmatrix},
\end{equation}
where $2\Delta = \gamma_2 - \gamma_1$. This analysis has reduced a generic two dimensional unitary optical scatterer to its minimum number of measurable parameters, two. $r$ controls the energy division and $\Delta$ determines the degree of non-reciprocity. If the device is reciprocal, $\Delta = 0$, and the only measurable parameter is the splitting ratio. In fact, for this two-dimensional case, enforcing $U$ to be gauge-invariant with respect to swap of labels $1 \leftrightarrow 2$ in turn implies that $U$ is a reciprocal matrix. Consequently, in two dimensions reciprocity and the only nontrivial gauge symmetry imply one another. This is intuitively rooted in the fact that reciprocity asserts invariance under permutation of source and detector, and the gauge symmetry $P$, which permutes the only two ports, effectively asserts the same thing.

Returning to Eq. (\ref{eq:reduced}) and setting $\Delta \rightarrow 0$, we arrive at the reciprocal two-port splitter with reflection probability $r^2$:
\begin{equation}
    U = 
\begin{pmatrix}
    -r & (1-r^2)^{1/2}\\
    (1-r^2)^{1/2} & r 
\end{pmatrix}. 
\end{equation}
This matrix is realized by a plate optic excited at normal incidence, with $r$ governed by the interface's refractive indices through the Fresnel equations. The device generates symmetric states when $r$ is set to $1/\sqrt{2}$ artificially; no gauge symmetry can induce this naturally. If another external phase gauge transformation is made, different sign conventions are obtained. For instance, to obtain a geometrically gauge-invariant form of this matrix, add additional external phases $\phi_1' = 3\pi/4$ and $\phi_2' = 5\pi/4$, producing 
\begin{equation}
    U' = 
\begin{pmatrix}
    -r e^{i3\pi/2} & (1-r^2)^{1/2} e^{i8\pi/4}\\
    (1-r^2)^{1/2} e^{i8\pi/4} & r e^{i5\pi/2}
\end{pmatrix}
=
\begin{pmatrix}
    i r & (1-r^2)^{1/2}\\
    (1-r^2)^{1/2} & i r 
\end{pmatrix}
\sim U.
\end{equation}

\subsection{Gauge-invariant \textit{U}(3) scatterers}

\begin{figure}[ht]
    \centering
    \includegraphics[width=0.6\linewidth]{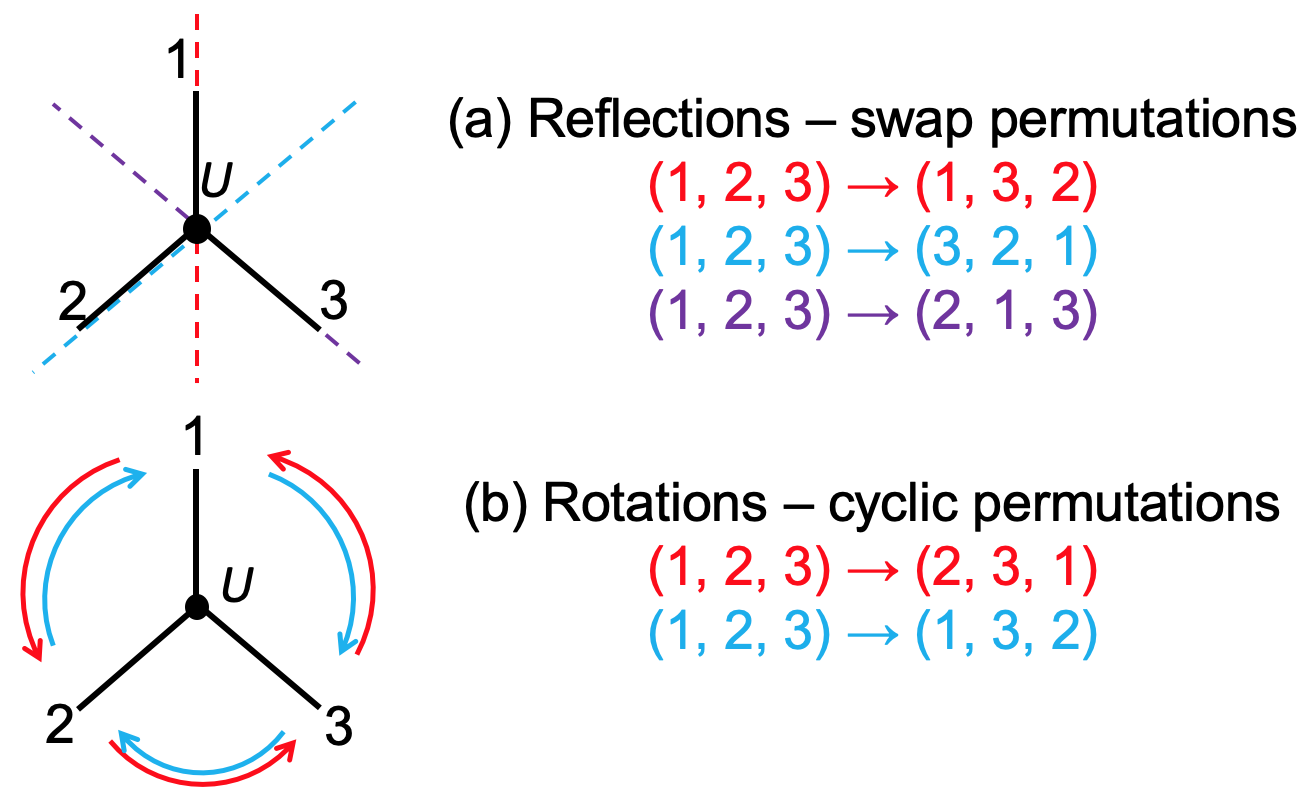}
    \caption{A generic three-dimensional scattering matrix $U$ can be equipped with two distinct gauge symmetries: one corresponding to a reflection symmetry (a) and one corresponding to rotation (b).}
    \label{fig:U3}
\end{figure}

In the two-dimensional case, there was only one gauge invariant transformation to consider, which was generated by swapping labels 1 and 2. In three dimensions, there are $3! = 6$ permutations comprising two nontrivial classes. These two classes are shown in Fig. \ref{fig:U3}. In this section, we present these gauge-invariant classes, highlighting how the gauge freedoms reduce the number of free variables in the scattering process description.

\subsubsection{Generalized three-port circulators}

In addition to the permutations that correspond to exchanging the labels of any given pair of modes, there are now distinct cyclic permutations: $(1, 2, 3)\rightarrow (3, 1, 2)$, with the associated permutation matrix
\begin{equation}\label{eq:cyclic}
    P = 
\begin{pmatrix}
0 & 0 & 1\\
1 & 0 & 0\\
0 & 1 & 0
\end{pmatrix}, 
\end{equation}
and $(1, 2, 3)\rightarrow (2, 3, 1)$, which is similarly associated with $P^T$. 

The general result of imposing a gauge symmetry on a scattering device $U$ will set various elements of $U$ equal to one another, reflecting the fact that input to some ports behaves no differently than the same input to others. For 
\begin{equation}
    U = 
\begin{pmatrix}\label{eq:U3}
    u_{11} & u_{12} & u_{13}\\
    u_{21} & u_{22} & u_{23}\\
    u_{31} & u_{23} & u_{33}
\end{pmatrix}
\end{equation}
enforcing $PUP^T = U$ generates the constraints $u_{11} = u_{22} = u_{33} \coloneqq a$, $u_{21} = u_{13} = u_{32} \coloneqq b$, and $u_{23} = u_{31}$ = $u_{12} \coloneqq c$, reducing $U$ into the form 
\begin{equation}\label{eq:U3cyc}
U = 
\begin{pmatrix}
    a & c & b\\
    b & a & c\\
    c & b & a
\end{pmatrix}.
\end{equation}
Each column of $U$ is a cyclic permutation of its neighbor. Matrices with this general property are known as circulant. The same form can alternatively be derived by assuming \textit{two} reflection symmetries: two exchanges is equivalent to a cyclic permutation in this space. For instance, $(1, 2, 3) \rightarrow (1, 3, 2)$ after one exchange and then after the second, $(1, 3, 2) \rightarrow (3, 1, 2)$ or $(1, 3, 2) \rightarrow (2, 3, 1)$. The net transformation in either case is a cyclic permutation of $(1, 2, 3)$. 

When $a = 0$, exactly one of either $b$ or $c$ must also be zero, and then the system (\ref{eq:U3cyc}) reduces to the circulator, Eq. (\ref{eq:circ}), or a matrix which is equivalent to it. This example illustrates that unlike in the two-dimensional case, a geometric gauge symmetry does not always induce reciprocal symmetry. Imposing reciprocity on $U$ in addition to its rotational symmetry would set $b = c$.

\subsubsection{Y couplers}

The reflection symmetry derived from the exchange of labels 2 and 3 produces a permutation matrix
\begin{equation}
    P = 
\begin{pmatrix}
    1 & 0 & 0\\
    0 & 0 & 1\\
    0 & 1 & 0
\end{pmatrix}.
\end{equation}
Imposing this symmetry on the general $U$ of Eq. (\ref{eq:U3}) constrain $U$ to be of the form
\begin{equation}
    U = 
\begin{pmatrix}
    a & c & c\\
    b & d & h\\
    b & h & d
\end{pmatrix}.
\end{equation}
This is the general form of a non-reciprocal Y coupler or Y branch. The $\mathbb{C}$-valued $a, b, c, d$ and $h$ would produce ten unconstrained degrees of freedom, but unitarity reduces this. Normalization lowers the count to eight, then the orthogonality conditions lower this count to five. Finally, the symmetry-preserving external phases $\phi_1$ and $\phi_2 = \phi_3$ can remove two more degrees of freedom, making the final count three.

\begin{figure}[ht]
    \centering
    \includegraphics[width=.6\linewidth]{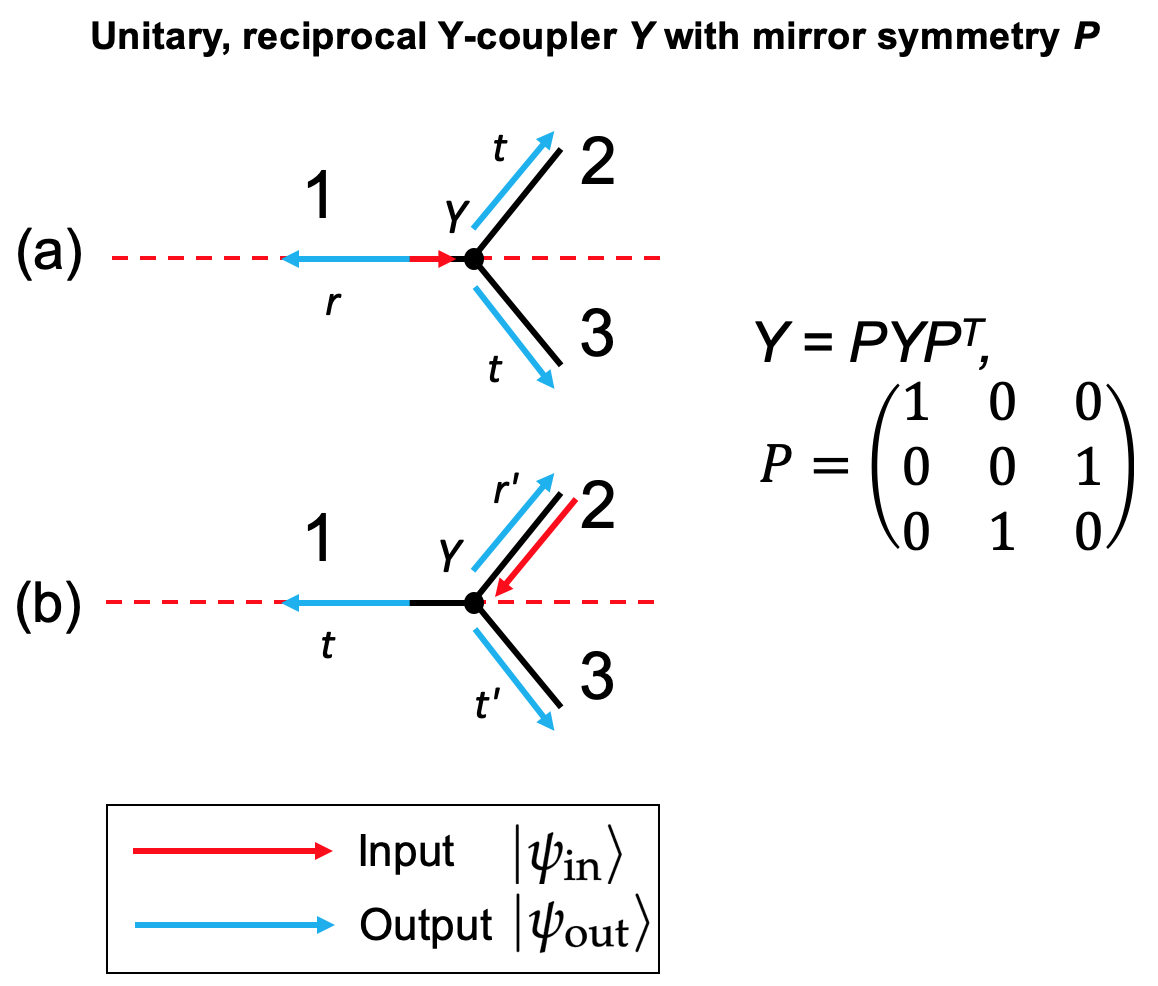}
    \caption{A Y coupler or Y branch is a three-port scattering device with a mirror symmetry, such as $P$ shown above. If the device is also reciprocal, its scattering matrix can only have a single degree of freedom $r$, taking on the form in Eq. (\ref{eq:Y}). As shown in diagram (a), this device always produces a mirror-symmetric distribution of energy when a photon enters port 1. The transmission coefficient $t = \sqrt{1-r^2}/2$. However, diagram (b) illustrates that input to port 2 or 3 generally does not produce another symmetric output state, as $r' = -(1+r)/2$ and $t' = (1-r)/2$. However, when $r$ is brought to zero, these amplitudes become $-1/2$ and $1/2$ respectively, so that their probabilities are equal, and the output state is symmetric. This form of the Y coupler exhibits added symmetry and is uniquely specified by the scattering matrix (\ref{eq:Ysymm}).}
    \label{fig:Y}
\end{figure}

If this gauge symmetry is also assumed with reciprocity, the added condition $b = c$ removes two more degrees of freedom, leaving just one parameter left. This case was studied in Appendix I of Ref. \cite{PhysRevA.110.023527}. The device, which is a back-reflecting symmetric Y-branch, was found to have the scattering matrix 
\begin{equation}\label{eq:Y}
    Y = 
\begin{pmatrix}
    r & \sqrt{(1 - r^2)/2} & \sqrt{(1 - r^2)/2}\\
    \sqrt{(1 - r^2)/2} & -(1 + r)/2 & (1 - r)/2 \\ 
    \sqrt{(1 - r^2)/2} & (1 - r)/2 & -(1 + r)/2
\end{pmatrix}, 
\end{equation}
where $r$ can freely be chosen in the set $(-1, 1)$. The different scattering states produced by this device, for single-photon input on precisely one of its ports, is shown in Fig. \ref{fig:Y}. For any value of $r$, the output state produced when a single photon enters port 1 will have a symmetric energy distribution under the same symmetry operator, $P$. And although the mirror symmetry ensures the behavior for light entering port 2 is the same as when it enters for port 3, the output state produced when one of these ports is excited is not necessarily symmetric in its energy distribution. This only occurs when $r$ is brought to zero, so that matrix becomes
\begin{equation}\label{eq:Ysymm}
    Y = 
\begin{pmatrix}
    0 & 1/\sqrt{2} & 1/\sqrt{2}\\
    1/\sqrt{2} & -1/2 & 1/2 \\
    1/\sqrt{2} & 1/2 & -1/2 
\end{pmatrix}. 
\end{equation}
This matrix, which originated from a nine-parameter unitary space, is uniquely 
specified by its various symmetries. This is often the intended scattering matrix when an otherwise unspecified Y branch is discussed. It is conventionally used as a compact beam-splitter to realize the transformation $a_1^\dagger \rightarrow (a_2^\dagger + a_3^\dagger)/\sqrt{2}$. The device is a hybrid in that it is feed-forward for input to port 1 but back-reflecting for its other inputs. The general behavior of this symmetric version of the Y coupler is shown in Fig. \ref{fig:comparison01} (c). This hybrid behavior can only occur in dimensions $d \geq 3$. The complete family of reciprocal, hybrid scattering devices in $U(3)$ is also derived in Appendix I of Ref. \cite{PhysRevA.110.023527}.

Y-couplers in practice can exhibit losses, violating the assumption of unitarity. This is especially common in integrated photonic implementations, where the light tends to radiate out of the device rather than back-reflect at ports 2 and 3. This is \textit{not} due to inferior design methodology or fabrication quality. Instead, the devices are only designed for the forward transformation. Optimizing the design for the reverse transformation while imposing the reflection gauge symmetry on the device structure would produce a Y coupler more similar to that of Eq. \ref{eq:Ysymm}. It would also automatically produce the forward transformation via reciprocity. For some applications, the back-reflections are viewed as an unwanted nuisance, so employing only the forward transform is useful in those cases. Nevertheless, to obtain an ideal Y-coupler, a simple decomposition with a beam-splitter can be used. This is shown in Fig. \ref{fig:decomposition1}, and can be employed in any platform, e.g., free-space optics, fiber optics, and integrated photonic waveguides.

\subsection{Gauge-invariant \textit{U}(4) scatterers}
\subsubsection{Beam splitters}

The general definition of an ideal beam-splitter is a $U(4)$ scattering device which possesses the block feed-forward gauge symmetry described above. In this case the relevant block feed-forward permutation invariance is with respect to the swap of the \textit{ordered pairs} $(1, 2) \longleftrightarrow (3, 4)$. This is associated with a diagonal reflection symmetry in traditional schematics of beam-splitters, as shown as the red dashed line in Fig. \ref{fig:bs}. The assumed symmetry produces the beam splitter scattering matrix
\begin{equation}\label{eq:BS}
B = 
\begin{pmatrix}
    0 & 0 & r_1 & t_2\\
    0 & 0 & t_1 & r_2\\
    r_1 & t_2 & 0 & 0\\
    t_1 & r_2 & 0 & 0
\end{pmatrix}
\end{equation}
reducing the device to an embedding a $U(2)$ scattering device in a four-dimensional space. Therefore, in essence, everything pertaining to the four-port optical beam-splitter is described in Sec. \ref{sec:U2} on two-dimensional scattering devices. 

The matrix $B$ in Eq. (\ref{eq:BS}) is not necessarily reciprocal. However, in this system, reciprocity is tied to a second reflection symmetry, over the anti-diagonal line in Fig. \ref{fig:bs}. The underlying permutation matrix is 
\begin{equation}
P = 
\begin{pmatrix}
    \sigma & 0\\
    0 & \sigma
\end{pmatrix},
\end{equation}
where $\sigma$ represents the swap operator for two modes. Once the block feed-forward symmetry is assumed, this symmetry $P$ is both a necessary and sufficient condition for reciprocity. If $B$ is reciprocal, adding external phases allows it to be put into the form
\begin{equation}\label{eq:BSr2}
B = 
\begin{pmatrix}
    0 & 0 & r & t\\
    0 & 0 & t & r\\
    r & t & 0 & 0\\
    t & r & 0 & 0
\end{pmatrix}. 
\end{equation}
\begin{figure}[ht]
    \centering
    \includegraphics[width=0.6\linewidth]{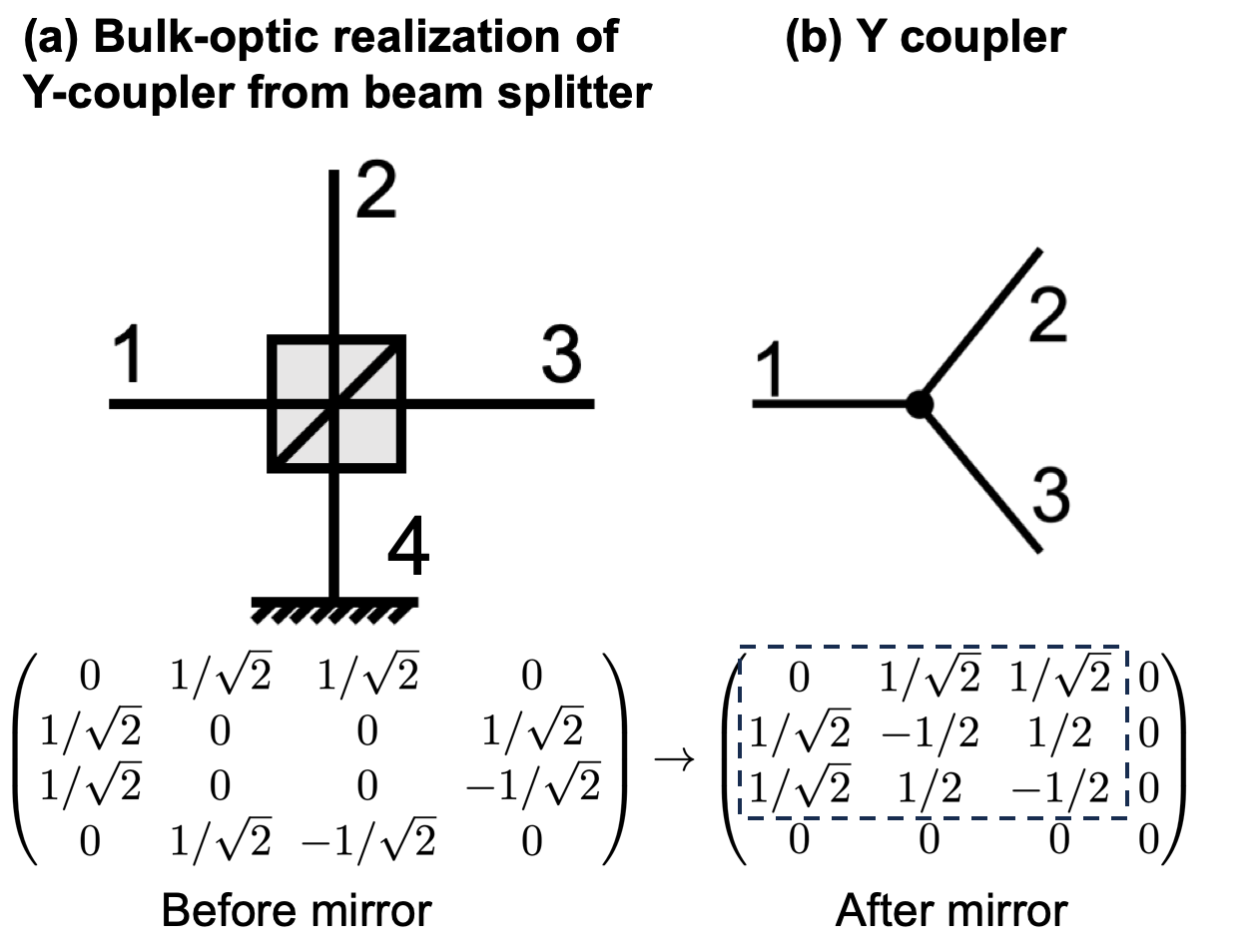}
    \caption{An ideal 50:50 beam-splitter and mirror (a) can produce an ideal symmetric Y-coupler (b). By labeling the beam-splitter as shown, its scattering matrix takes on the form below. The first row and column already match those for the intended Y coupler. However, light entering port 2 or 3 loses energy to port 4. Placing the mirror there redirects the energy back into the beam splitter, splitting the beam a second time. This ultimately produces the Y coupler matrix. The phase acquired between the mirror and beam-splitter is immaterial, since external phases can always be used to equivalently adjust its value.}
    \label{fig:decomposition1}
\end{figure}

\begin{figure}
    \centering
    \includegraphics[width=0.4\linewidth]{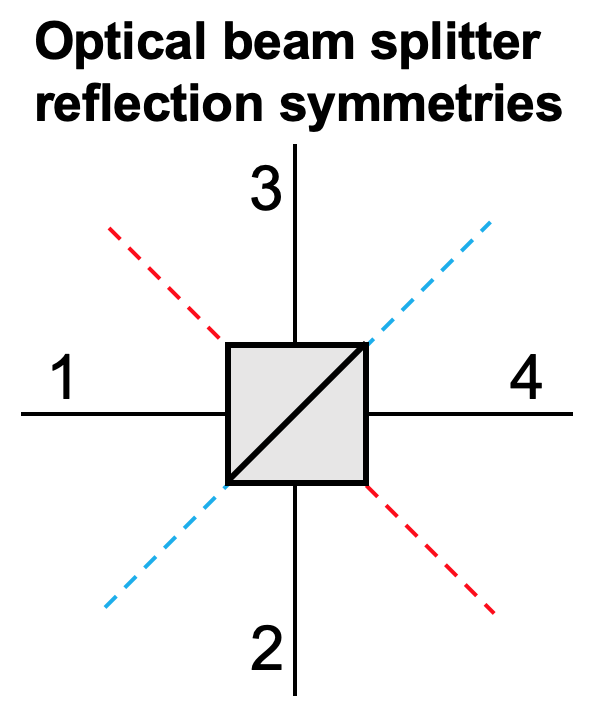}
    \caption{An optical beam-splitter is defined as any four-port scattering device which satisfies the following conditions. First, it exhibits zero coupling between (1, 2) and itself as well as (3, 4) and itself. Second, it possesses the red line mirror symmetry, induced by scattering matrix invariance to the permutation of label pairs $(1, 2) \longleftrightarrow (3, 4)$. The symmetry over the blue axis is induced by invariance to the simultaneous swap of the mode labels $1 \leftrightarrow 2$ and $3 \leftrightarrow 4$. After the original (red) symmetry constraint is enforced, the second (blue) one becomes a necessary and sufficient condition for reciprocity.}
    \label{fig:bs}
\end{figure}

Loopless graphs comprised solely of beam-splitters will generally produce a higher-dimensional scattering devices which also possess block feed-forward gauge symmetry. These \textit{decompositions}, which represent a higher dimensional scattering device as a graph of lower dimensional ones, can be made to span the full unitary group. Two well-known decompositions with the underlying reflection symmetry of all input and output modes are due to Reck and Clements \textit{et al.} \cite{PhysRevLett.73.58, Clements:16, Kumar_2021, yasir2025compactifyinglinearopticalunitaries}. These both require $\mathcal{O}(d^2)$ beam-splitters and controllable phase shifts to span $U(d)$. These meshes are versatile, but in practice, they suffer from error accumulation and nontrivial reprogramming procedures. This has limited their size $d$ to about a few dozen modes today \cite{Taballione_2023, Maring2024}. A non-reciprocal phase placed inside one of these meshes would break the input/output mirror symmetry, leading to an overall non-reciprocal $U$. In terms of Eq. (\ref{eq:Ugen}), this scattering matrix would have $A = D = 0$ but $B \neq C$. 

\subsubsection{Grover four-ports}
Although a reciprocal, balanced beam-splitter commutes with the block feed-forward gauge symmetry operator, there are a number of other potential gauge symmetries that it lacks. However, \textit{the Grover four-port, }
\begin{equation}
G = \frac12
\begin{pmatrix}
-1 & 1 & 1 & 1\\
1 & -1 & 1 & 1\\
1 & 1 & -1 & 1\\
1 & 1 & 1 & -1
\end{pmatrix}
\end{equation}
\textit{is unchanged by all possible permutations of its labels, and thus commutes with any gauge symmetry operator. Beyond that, it produces a fully-uniform, symmetric energy distribution when any one of its ports is excited. It is precisely the uniform energy distribution which is invariant to any permutation of mode labels.}

The matrix originally appeared as a quantum logic gate in the Grover search algorithm \cite{Grover_1997}. It has since been employed as a general scattering coin for quantum walks \cite{Kempe_2003} and now more recently as an optical scattering device. Having the same number of ports as a beam-splitter allows the Grover coin to replace the latter in any system, effectively boosting the underlying dimensionality. These two devices are compared in Fig. \ref{fig:comparison01}. 

A general $d$-dimensional Grover coin can be defined by first enforcing reciprocity, a reflection symmetry, and a rotational symmetry, producing a reciprocal, circulant matrix, and then by minimizing the reflection probability \cite{PhysRevA.110.023527}. This produces a reflection amplitude $r = 2/d-1$ with a tranmission amplitude is $t = 2/d$. In general, Grover coins in one dimension can easily be converted to those in the other dimensions, by either closing off ports with a mirror, or by connecting two coins together \cite{PhysRevA.107.052615}. General circulant matrices have been studied for quantum walks \cite{PhysRevA.106.022402}. 

\begin{figure}[ht]
    \centering
    \includegraphics[width=0.5\linewidth]{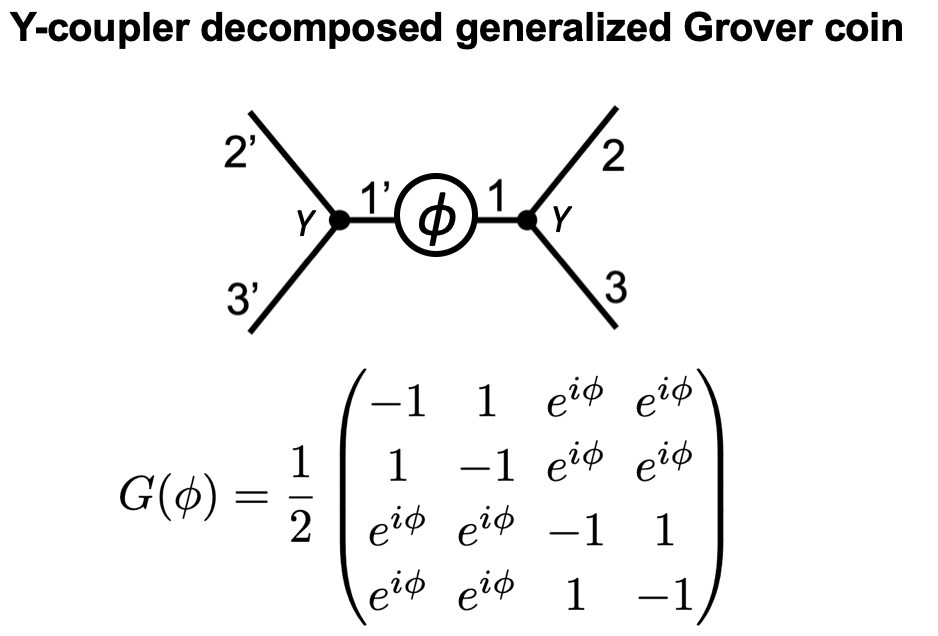}
    \caption{Forming a generalized four-port Grover coin using two symmetric Y couplers. The standard Grover coin is obtained when $\phi = 0$.}
    \label{fig:decomposition2}
\end{figure}

Realization of a four-port Grover coin is important for interferometric applications which are discussed in the next section. Although a static, single-element Grover coin has yet to be experimentally realized, a Grover coin can and has been obtained using a decomposition of two Y-couplers \cite{PhysRevA.110.023527, Schwarze:24}. The Y-couplers themselves can be decomposed with a beam-splitter and a mirror, as shown in Fig. \ref{fig:decomposition1}. The Grover decomposition is shown in Fig. \ref{fig:decomposition2}. It is formed by connecting the feed-forward ports of two symmetric Y-couplers to each other. This is a slightly more general decomposition, due to the bridge phase $\phi$. $\phi$ parametrizes a family of uniform-probability $U(4)$ scattering devices. These various static devices produced are generally not gauge-equivalent to one another in the external phase sense. The standard Grover coin occurs for $\phi = 0$.

\section{Optical interferometry}\label{sec:interferometry}
The previous section dealt entirely with static or passive scattering devices. These are by definition presumed to be fixed for excitations of a given frequency and polarization mode. Static scatterers are generally represented as a graphical node, like the Y coupler shown in Fig. \ref{fig:Y}.

This present section will consider ways in which multiple passive nodes can form a graph, called an optical \textit{interferometer}, which is commonly used as a central element in optical sensors. This graph will generally be both a function of the scattering matrices forming it, as well as the phases acquired on the edges between connected nodes. Controllable phase shifters could also be equivalently affixed to an unweighted graph as block feed-forward nodes of their own. In either case, when multiple scattering devices are treated as part of the same system but are left unconnected, a block-diagonal aggregate scattering matrix is obtained. Adding connections couples energy between the constituent scattering nodes, breaking the block-diagonal structure. Any remaining unconnected edges allow light to couple in and out of the network. Thus, the interferometer itself is another scattering device which is tunable or active. 

A photon entering an open or unconnected port of a general interferometer can be viewed as conducting a random walk on its graph. This variant of random walk is known as a quantum walk, since the paths it traverses are able to coherently interfere; this distinction is embodied in the use of a unitary time-evolution operator instead of the traditional probability matrix used in studies of Markov chains. The study of quantum walks in general forms a subdomain of general quantum information processing. It has been shown that universal computation and efficient search on a graph can be achieved with quantum walks \cite{PhysRevLett.102.180501, Lovett_2010, PhysRevA.70.022314, PhysRevLett.129.160502, KENDON2008187}. For an introduction to the field, consult Ref. \cite{Kempe_2003}.

For each set of input phases and constituent scattering devices, the action of any interferometer can be computed by finding the steady-state distribution of the walk, which is the outputted scattering state in the limit of long time. This can be found numerically using a finite-element assembly approach. This involves using the graph interconnect information to construct a global time-evolution operator for the walk, enforcing boundary conditions pertaining to the unconnected ports, and finally computing its action in the long-time limit. This is effectively accomplished by finding the eigenspectrum of the time-evolution operator, either directly, or through a matrix inversion problem. The general approach is described in Ref. \cite{Schwarze:242}. For the special case of a chain of two-dimensional scattering devices, this finite-element method has been shown to reduce to the Redheffer star product which is extensively used to compute the aggregate scattering matrix of two or more coupled optical thin films \cite{Redheffer}.

The quantum walk model of an interferometer typically begins with the assumption that the input single-photon state is perfectly coherent, which is akin to assuming the source is perfectly monochromatic. This in turn permits the summation of paths with arbitrarily large optical path length differences to obtain the output after interference. For a real source, interference will only be observed among paths whose corresponding path lengths are smaller than the coherence length of the source. The other paths will only appear in the output as an incoherent background. Adoption of a very narrow linewidth source can extend the coherence length to several kilometers or more, making this assumption valid to an excellent approximation. This discussion holds in an analgous way for non-planar sources with finite spatial coherence. To obtain the most ideal behavior, a high degree of collimation and alignment is necessary.

The general methodology of the quantum walk finite-element approach can also be followed algebraically, for the sake of finding an interferometer's scattering matrix as a function of its parameters. A common obstruction, however, is arriving at a closed-form expression for the inverse or eigenspectrum of the constructed $d \times d$ \textit{matrix function}. However, this can often be accomplished in low-dimensional systems; many examples will be discussed in the sections to follow. We will first review the operating principles of several traditional interferometers. Then, by substituting their components for higher-dimensional gauge-symmetric scattering devices, we show these devices gain access to new degrees of freedom. These in turn lead to various enhancements of the phase readout, thus resulting in a substantial improvement in the resolution of optical sensors.

\subsection{Interferometric gauge freedoms}

The gauge freedoms of passive scatterers naturally generalize to active ones. External phases can be applied to open ports in the same way as before, and phase shifts can also be applied to the ports of the internal nodes. This shifts the total phase acquired between the scattering nodes by a constant. Because the true global value of these phases are indeterminate, the translation by another constant will not change the physical situation either. It can be freely absorbed into or removed from the definition of the acquired phase between two nodes. Therefore, two interferometers are considered equivalent in this sense if a suitable translation of each internal and external phase of either system produces the other system, \textit{for all values of the internal phases}. In other words, it is obtained when the parameter spaces of the two networks coalesce when the input parameters are shifted by appropriate constants. 

This definition is employed so that changes in the external phase convention of the \textit{internal} scattering nodes (such as Hadamard vs. symmetric beam splitter) do not affect the equivalence class of the overall network. It is a different comparison than that for two static scattering devices, where only external phases can be placed. To highlight the distinction in an example, consider the generalized Grover coin decomposition of Fig. \ref{fig:decomposition1} and let the scattering matrix of the network be denoted $G_1(\phi)$. The standard Grover coin is obtained as a static device when $\phi$ is fixed at the value 0. For $\phi \neq 0$, the static device produced is in general not equivalent to the standard Grover coin. However, now form a copy of this device, $G_2$, and add a second phase in the bridge, $\phi_b$. $\phi_b$ may or may not be coming from external phases placed at port $1$ and/or port $1'$, since that distinction is immaterial. Nevertheless, the total bridge phase is now $\phi + \phi_b$. Clearly for nonzero $\phi_b$ the \textit{static} devices produced for fixed values of $\phi$ and $\phi_b$, $G_1(\phi)$ and $G_2(\phi + \phi_b)$ will not be equivalent. But $\phi$ cannot be defined up to an absolute, so it can be shifted freely. Redefining the phase $\phi \rightarrow \phi - \phi_b$ in $G_2$ causes $G_1$ and $G_2$ to be equal for all values of $\phi$, signifying the equivalence of the two networks. This example is trivial because the compared networks have the same graph topology. In some more interesting cases, equivalence in this sense can be established between different graphs. 

Geometric gauge symmetry of a scattering node is also able to produce a symmetrical scattering graph, assuming the graph itself preserves the symmetry in its interconnect configuration. Since an external port relabeling of the graph is equivalent to a geometric transformation enacted on it, the entire graph can remain invariant to such a transformation if the underlying vertices and phase parameters maintain the symmetry. One practical use of interferometric gauge symmetry is to simplify the computation of the network output: if symmetry implies certain sets of input ports all scatter light in the same way, only one input case has to be studied, and then the rest follow by suitable permutations of the graph parameters. Gauge-invariant interferometric graphs are readily constructed in general: first form a graph which is invariant to some permutation to any subset of its nodes. This can correspond to a global symmetry of the overall network topology, such as an invariance to rotations of the whole graph by a fixed angle. Then, assign the same scattering matrix to each relevant node, providing them the same vertex label or color.
\begin{figure}[ht]
    \centering
    \includegraphics[width=0.75\linewidth]{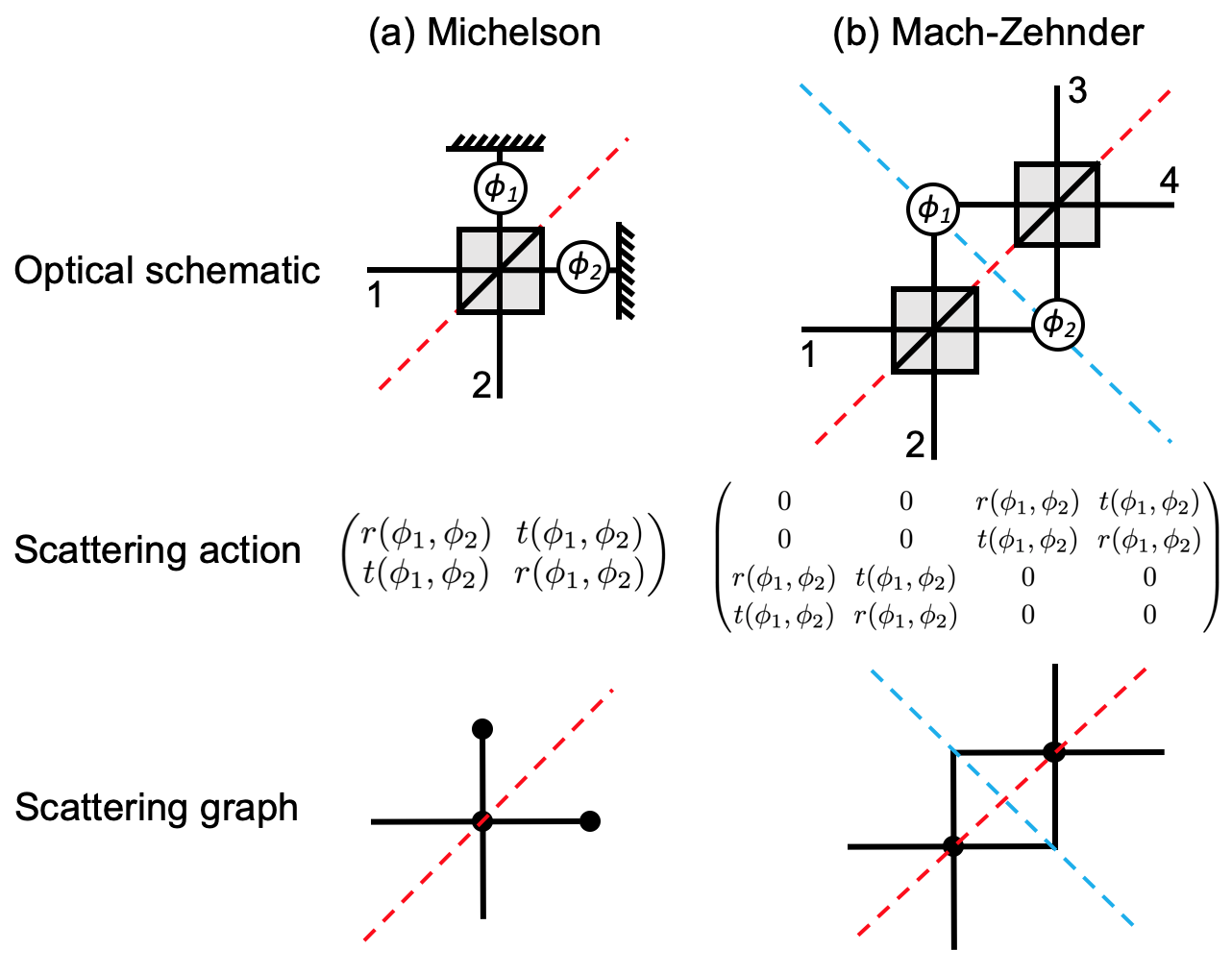}
    \caption{Two common interferometric configurations, due to Michelson (a) and Mach and Zehnder (b). Like static scattering devices, the open ports of interferometric graphs can also be freely relabeled, naturally extending the notion of geometric gauge symmetry to scattering graphs. Assuming the underlying 50:50 beam-splitters and phase shifters are reciprocal, the devices will possess the reflection symmetry indicated by the red dashed line. This symmetry implies that input to any port on one side of the line will behave the same as input to its mirrored counterpart, so long as $\phi_1$ and $\phi_2$ are exchanged. The Mach-Zehnder interferometer can be viewed as constructed by mirroring the Michelson interferometer over the blue dashed line. This line corresponds to the device's block feed-forward symmetry. Since these devices inherit the symmetries and overall dimensionality of their lower dimensional counterparts (c.f. Fig. \ref{fig:bs}), they are often viewed as tunable versions of them. That is, a Michelson interferometer can be viewed as a phase tunable plate splitter while the Mach-Zehnder interferometer can be viewed as a tunable beam splitter. These active building blocks can replace their passive equivalent in any larger interferometric graph. The phase gauge freedom also allows the placement of external phases at ports 1 and 2 and translation of the values of $\phi_1$ and $\phi_2$ by arbitrary fixed constants. These can be used to put the overall scattering matrix for each device in the form shown beneath its schematic. Here the correspondence between nodes and local scattering matrices is clear, since the associated optical schematic is shown with it. In general, the nodes of a scattering graph need to be clearly associated to a specific scattering matrix, e.g. through a vertex label or color.}
    \label{fig:mimzi}
\end{figure}

Two well-known interferometric configurations, the Michelson and Mach-Zehnder interferometers, are shown in Fig. \ref{fig:mimzi}. As functions of their internal phases $\phi_1$ and $\phi_2$, these interferometers tunably produce the complete space of reciprocal $U(2)$ and $U(4)$ scattering devices. They inherit the symmetries from the reciprocal beam splitters which form them. In the case of the Michelson (\ref{fig:mimzi}(a)), the exchange of labels $1 \leftrightarrow 2$ of the external ports \textit{and the phase parameters} leaves the scattering response unchanged. Thus, the computed output corresponding to an input excitation at port 1 can be converted to the output corresponding to input at port 2 simply by exchanging the output mode labels and exchanging the phase parameter labels. The Mach-Zehnder interferometer shown in Fig. \ref{fig:mimzi}(b) can be generated by mirroring the Michelson device about the blue dashed line. This preserves the symmetry defined by the red dashed line. The blue dashed line symmetry represents the gauge invariance tied to the block feed-forward symmetry $(1, 2) \longleftrightarrow (3, 4)$ of the configuration. Together, these two symmetry lines allow the full device output to be found immediately from just one calculation corresponding to input at any single port.

The general output calculation entails starting with an initial state which is a single photon impinging a given port, such as $|\psi_{\text{in}}\rangle = a_j^\dagger|0\rangle$. Then the scattering matrices are used to sequentially propagate the state though the graph until every state component has reached the output. This is equivalent to finding all the possible paths the photon can travel from an input $j$ to each output port $i$. Each individual path is first weighed by the product of all the scattering amplitudes it encounters going from port $j$ to $i$. The final scattering matrix element is found from the sum of these weighted paths, corresponding to coherent interference of all the paths. The general problem of path counting can grow unwieldy, especially as resonant structures are introduced to the graph, causing the number of possible paths to grow indefinitely. Some cases can be explicitly summed, but other times an automated path-counting and summing algorithm is needed,  which is what Ref. \cite{Schwarze:242} accomplishes. 

For both of the devices in Fig. \ref{fig:mimzi}, there are always only two possible paths, one for each arm $j$ containing the phase $\phi_j$. Their only difference in output is a meaningless global phase shift of $\pi$, originating from the common convention of adding a $\pi$ phase shift to the state when it reflects at a mirror. Whether that phase can be observed physically or not, it can be lumped into the definition of $\phi_1$ and $\phi_2$. This causes the device outputs to be identical. We will consider a photon inputted to port 1 of either device, and assume without loss of generality that the beam splitters take on the Hadamard convention of Eqs. (\ref{eq:H1}-\ref{eq:H2}). Then, there is an initial splitting into two state components, one in each arm. These each acquire a phase $\phi_j$ and then are reintroduced to a beam-splitter in some way, splitting a second time. The possible paths from port 1 to port 1 (3) are double-reflection with phase $\phi_1$ and double-transmission with phase $\phi_2$. The corresponding path-amplitudes are $(1/\sqrt{2})(e^{i\phi_1})(1/\sqrt{2}) = e^{i\phi_1}/2$ and similarly $e^{i\phi_2}/2$. The aggregate reflection amplitude for the device is then $r(\phi_1, \phi_2) = (e^{i\phi_1} + e^{i\phi_2})/2$ The transmission amplitude is similarly $t(\phi_1, \phi_2) = (e^{i\phi_1} - e^{i\phi_2})/2$, where the minus sign is incurred from the back-side beam-splitter reflection when using the Hadamard beam-splitter convention. 

\begin{figure}[ht]
    \centering
    \includegraphics[width=0.75\linewidth]{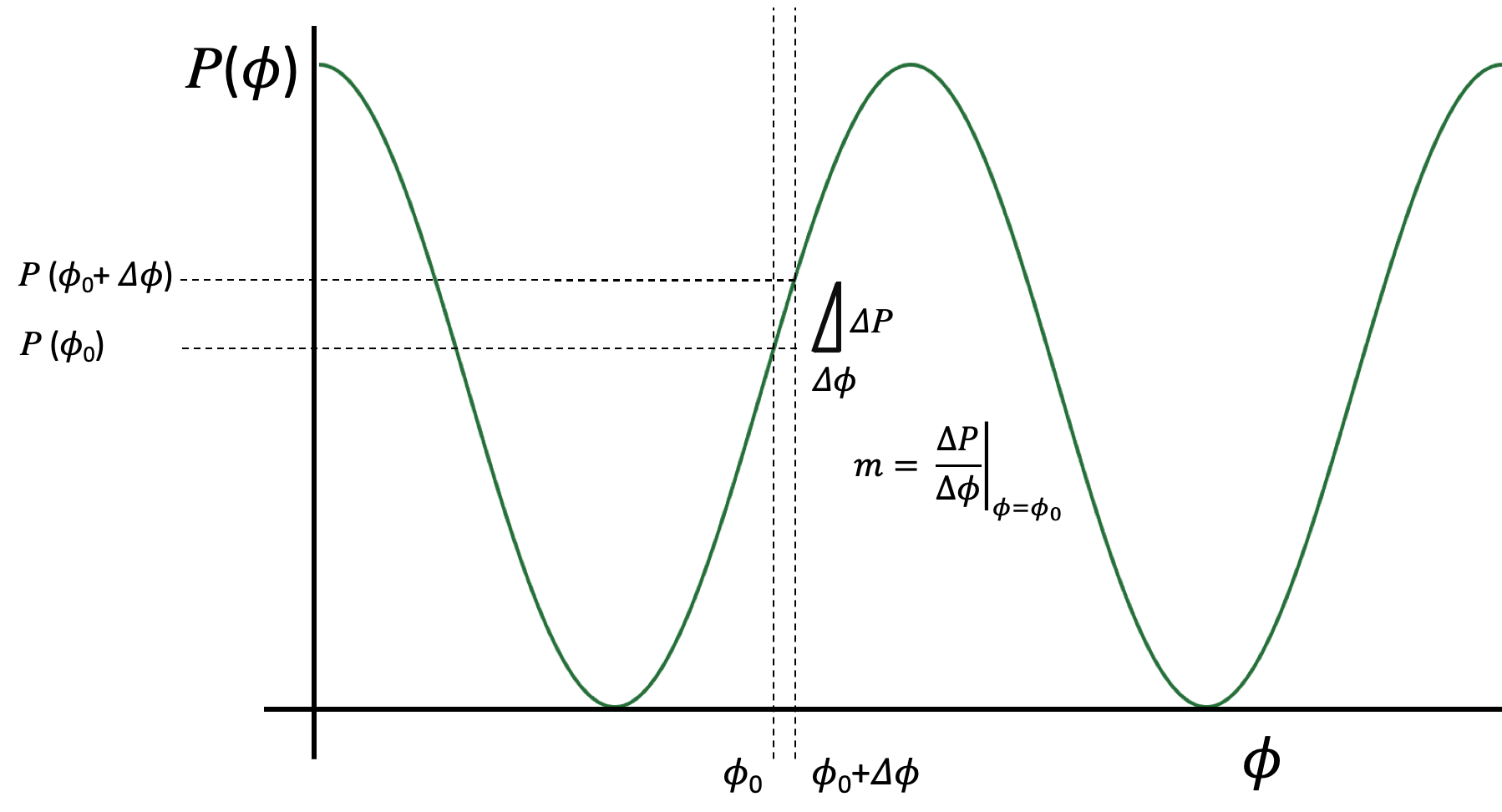}
    \caption{Sinusoidal interferogram produced by two-beam interference, such as with a Michelson or Mach-Zehnder interferometer. This basic interferometric configuration can be used as a phase readout. First the system is placed at a bias point $\phi_0$, verified by measuring the corresponding probability $P(\phi_0)$. Then, a small change in phase $\Delta \phi$ induces a change in probability $\Delta P$ given by the product of $\Delta \phi$ and the slope of the probability at the bias point. A larger slope $dP/d\phi$ at $\phi_0$ naturally produces a more sensitive response to phase changes, allowing smaller phase changes to be detected.}
    \label{fig:2beam}
\end{figure}

Thus the aggregate scattering matrix for any lossless Michelson interferometer is equivalent to
\begin{equation}\label{eq:MI}
U_{\text{MI}}(\phi_1, \phi_2) = 
\begin{pmatrix}
    r(\phi_1, \phi_2) & t(\phi_1, \phi_2)\\
    t(\phi_1, \phi_2) & r(\phi_1, \phi_2)
\end{pmatrix}
\end{equation}
while that for a Mach-Zehnder is equivalent to, in block form, 
\begin{equation}
U_{\text{MZI}}(\phi_1, \phi_2) = 
\begin{pmatrix}
    0 & U_{\text{MI}}(\phi_1, \phi_2)\\
    U_{\text{MI}}(\phi_1, \phi_2) & 0
\end{pmatrix}.
\end{equation}
The underlying scattering amplitudes are 
\begin{subequations}
\begin{align}
    r(\phi_1, \phi_2) &= \frac{e^{i\phi_1} + e^{i\phi_2}}{2} = e^{i\phi_1/2+i\phi_2/2} \cos(\phi_1/2 - \phi_2/2), \\
    t(\phi_1, \phi_2) &= \frac{e^{i\phi_1} - e^{i\phi_2}}{2} = e^{i\phi_1/2+i\phi_2/2} i\sin(\phi_1/2 - \phi_2/2).
\end{align}
\end{subequations}
from these forms it is easy to see that the scattering probability for reflection is $R = |r|^2 = \cos^2((\phi_1 - \phi_2)/2)$ while for transmission is $T = |t|^2 = \sin^2((\phi_1 - \phi_2)/2)$. This sinusoidal modulation, which is only sensitive to the optical phase difference $(\phi_1 - \phi_2)$, is characteristic of the two-beam superposition accomplished by devices such as these. A plot of either scattering probability $P$ versus either phase $\phi$ is known as an interferogram. $\phi$ itself could be viewed as a phase difference $\Delta\phi = \phi - \phi_g$ for an indeterminable global phase $\phi_g$. The typical sinusoidal interferogram for two-beam interference is shown in Fig. \ref{fig:2beam}. The system can act as a basic phase sensor: for small changes in the input phase about some bias point $\phi_0$, the output change in probability $\Delta P$ is given by the product of the input phase change and the slope at the bias point. A bias point with a larger slope will then have heightened sensitivity. 

\subsection{Traditional interferometric sensing systems}

\begin{figure}[ht]
    \centering
    \includegraphics[width=0.7\linewidth]{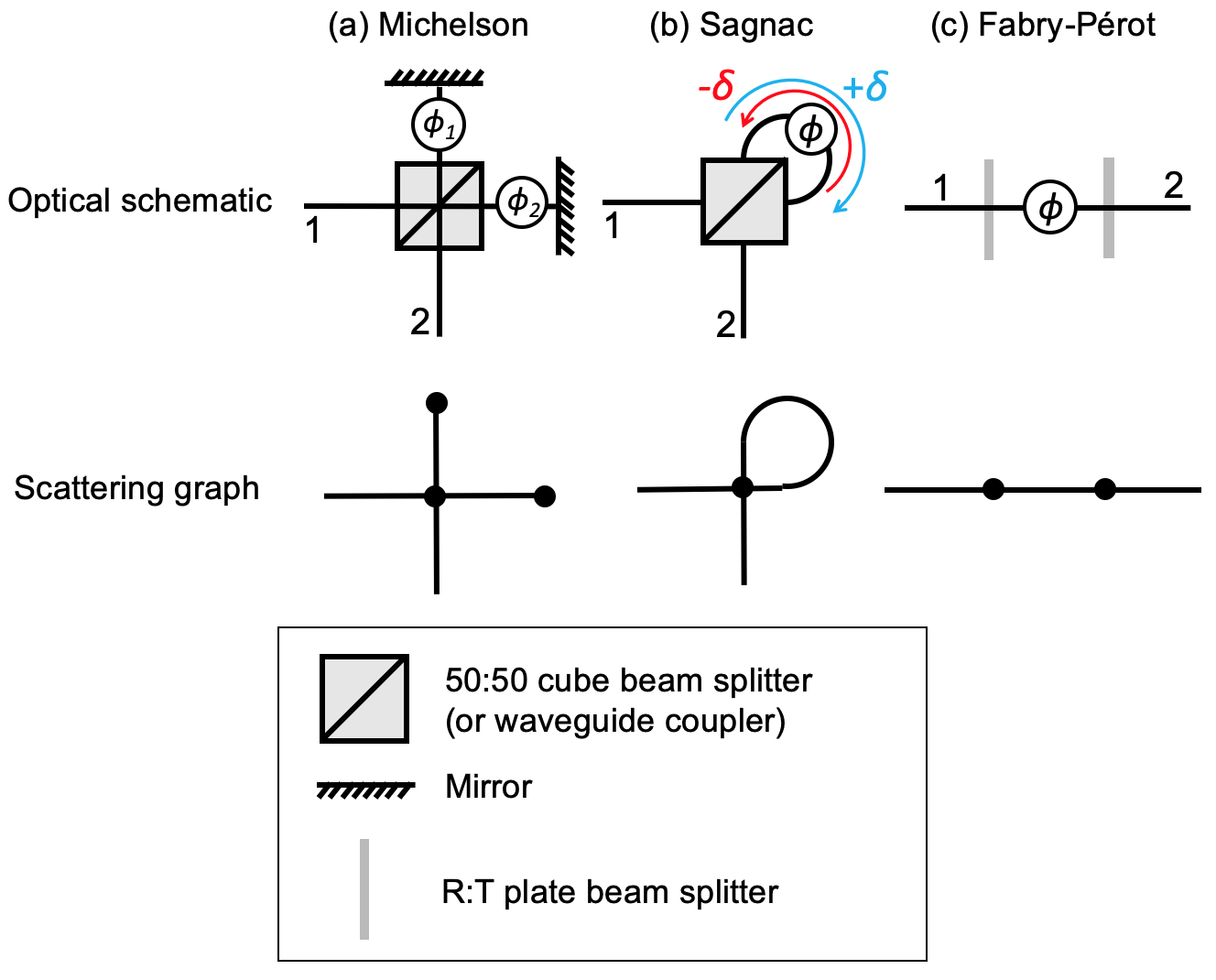}
    \caption{Traditional low-dimensional interferometric configurations. The optical schematic is shown above the associated scattering graph. The Michelson (a) is a common tool for inference of reciprocal phases, while the Sagnac (b) is used to infer non-reciprocal ones. The Fabry-Pérot interferometer (c) is a simple resonant cavity typically employed as a narrow-band filter of light.}
    \label{fig:traditional}
\end{figure}

Three two-port interferometers are shown in Fig. \ref{fig:traditional}. In case (a) is the Michelson configuration, which was discussed in the previous section, is the most simple phase sensor of \textit{reciprocal} phase shifts. Even if $\phi_1$ or $\phi_2$ were non-reciprocal, meaning the phase acquired depended on the propagation direction, this could not be observed in a Michelson interferometer. This is because the double-pass through each arm gives the summed phase for each direction as a single phase shift: $\phi_j = \phi_{jF} + \phi_{jB}$. 

The configuration of Fig. \ref{fig:traditional}(b) is the Sagnac interferometer, which is the simplest configuration for \textit{non-reciprocal} phase inference. By converting the two spatially-separated arms of a Michelson interferometer into a loop, any non-reciprocal difference in phase acquired going clockwise around the common-path loop versus counterclockwise will be evident. Let the total clockwise phase be $\phi + \delta$ and the total counterclockwise phase be $\phi - \delta$. Since this configuration is essentially the same two-beam interference present in a Michelson or Mach-Zehnder interferometer, those results can be applied here, but with these phases. The corresponding transmission probability is then
\begin{equation}
    T = |t|^2 = \sin^2(\phi_1/2 - \phi_2/2) = \sin^2((\phi + \delta)/2 - (\phi - \delta)/2) = \sin^2(\delta).
\end{equation}
Thus, the transmission probability is zero for a fully reciprocal phase ($\delta = 0$) and increases as the non-reciprocal component $\delta$ increases. However, the effect can be challenging to observe, due to the intrinsic weakness of non-reciprocal phase sources, combined with this weak $\mathcal{O}(\delta^2)$ response for small $\delta$. Nevertheless, it allows the non-reciprocal portion of a phase shift to be naturally isolated from the reciprocal one and extracted from a direct measurement. The configuration is named for Georges Sagnac, who established the existence of a physical effect by which rotation of a looped optical path induces a non-reciprocal phase shift \cite{RevModPhys.39.475}. The non-reciprocal phase shift $\delta$ is proportional to the product of the loop area and angular velocity and inversely proportional to the optical wavelength. It is the fundamental operating principle of optical gyroscopes, and was famously used in 1925 by Michelson and Gale to measure the rotation rate of the earth \cite{MICHELSON1925}. The last century has seen a number of applications and generalizations to the Sagnac effect proposed \cite{RevModPhys.57.61, 10.1063/1.522521, PhysRevLett.93.143901, shanavas2025nonlinearsymmetrybreakingenhance}.

The third device, in Fig. \ref{fig:traditional}(c) is the parallel-plate Fabry-Pérot interferometer.\footnote{If the plates are movable so that the central phase shift $\phi$ can be tuned, the device was traditionally called an interferometer, whereas if the plates were fixed, it was known as an etalon.} The multi-pass structure allows $\phi$ to accumulate many times, sharpening the phase response around certain values of $\phi$. This translates directly to a sharpened frequency or narrowband wavelength dependence about certain wavelength centers. 

The normally-incident transmission amplitude of the Fabry-Pérot interferometer is readily calculated by summing the different paths from input to output. For an input to port 1, each path is indexed by the number of bounces it takes before exiting at port 2. The principle beam transmits directly through both plates, acquiring an amplitude $t_1e^{i\phi}t_2$, where $t_j$ is the transmission coefficient of plate $j$ and $\phi$ is the phase acquired between the plates. The plates are assumed to be reciprocal $U(2)$ scattering devices. This allows the reflections coefficients to be written $r_j = i\sqrt{R_j}$ where $R_j$ is the plate reflection probability. Then, $t_j = \sqrt{1-R_j}$. 

\begin{figure}[ht]
    \centering
    \includegraphics[width=0.75\linewidth]{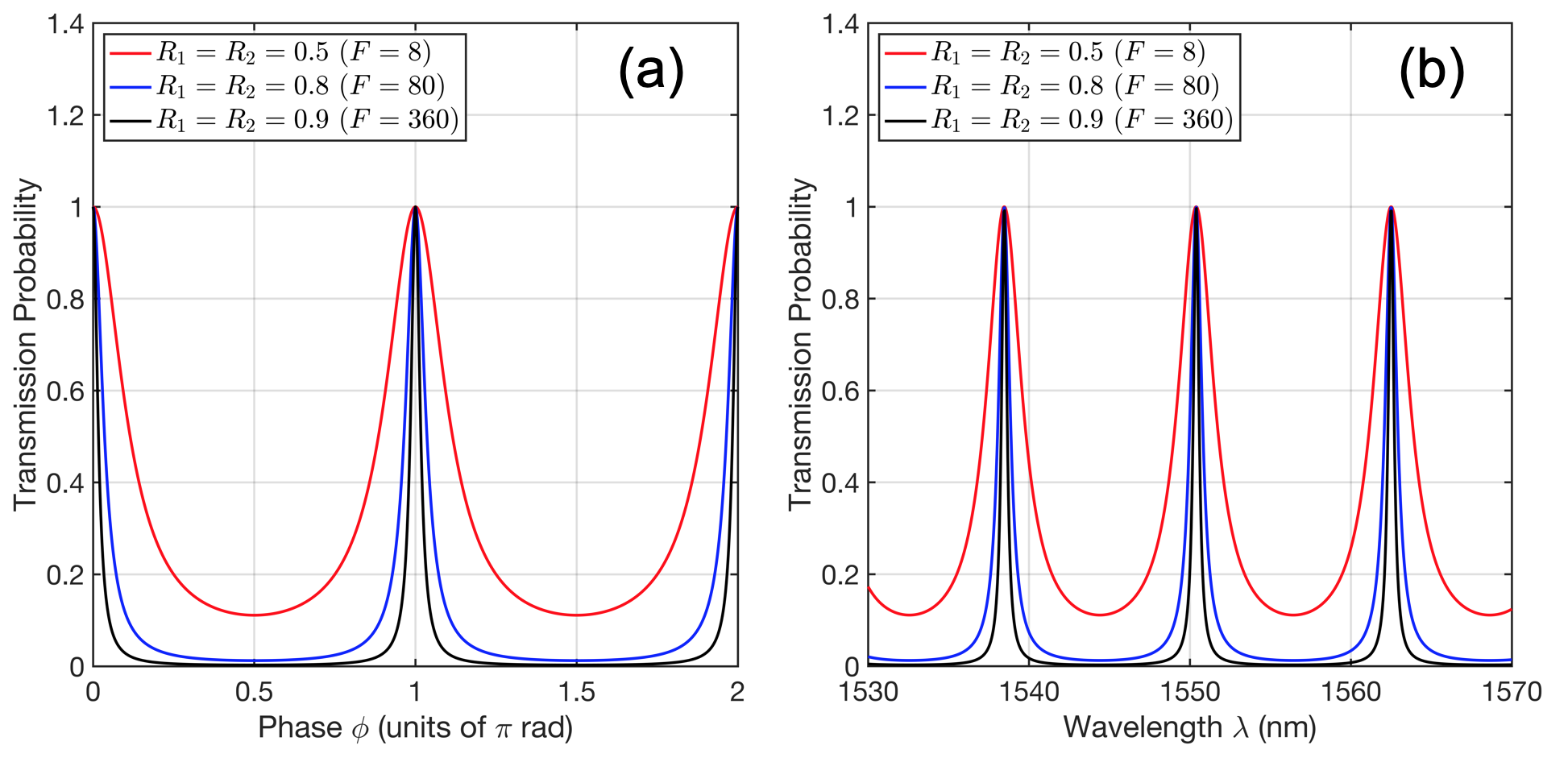}
    \caption{Transmission response of a symmetric Fabry-Pérot interferometer (Fig. \ref{fig:traditional}, (c)), vs. the half-round-trip phase $\phi$ (a) and wavelength (b). }
    \label{fig:fpi}
\end{figure}

The $n$th beam transmits through the first plate up to the second plate, acquiring $t_1 e^{i\phi}$, then makes $n$ bounces between the two plates, acquiring $(r_1r_2e^{2i\phi})^n$, and finally passes through the last plate with amplitude $t_2$. The net amplitude for the path conducting $n$ bounces before transmitting is thus $t_1t_2e^{i\phi}(r_1r_2)^n e^{(2i\phi)n}$. Summing these paths produces the Fabry-Pérot transmission amplitude, $t$, given by
\begin{equation}
    t = t_1t_2 e^{i\phi} \bigg (1 + \sum_{n=1}^\infty (r_1r_2 e^{i2\phi})^n \bigg) = t_1t_2 e^{i\phi} \bigg (1 + \sum_{n=0}^\infty (r_1r_2 e^{i2\phi})^n - 1\bigg).
\end{equation}
Evidently, this amplitude and the associated transmission probability $|t|^2$ can be decomposed into a product of the output which would have resulted without path interference, $|t_1|^2|t_2|^2$, and an interference factor in parentheses. Its sum can be evaluated as a geometric series, which converges whenever the product of plate reflections $|r_1r_2| = |r_1||r_2|$ is less than 1. This fails to occur only if either plate is fully reflecting, in which case no light will transmit. Otherwise, 
\begin{equation}
    t(\phi) = t_1t_2 e^{i\phi}\bigg (\sum_{n=0}^\infty (r_1r_2 e^{i2\phi})^n\bigg ) = \frac{t_1t_2e^{i\phi}}{1 - r_1r_2 e^{i2\phi}}.
\end{equation}
Using the fact that $r_j = i\sqrt{R_j}$ and $T_j = |t_j|^2 = 1-R_j$, the transmission probability $T = |t|^2$ can be written
\begin{equation}
    T(\phi) = \frac{(1 - R_1)(1-R_2)}{1 + R_1R_2 - 2 \sqrt{R_1R_2}\cos(2\phi)}.
\end{equation}
Now if the optical plates are symmetric, so that $R_1 = R_2 = R$, $T$ can be simplified using the fact that $\cos2\phi = 1-2\sin^2\phi$. Substituting this into the denominator yields
\begin{equation}
    1 + R^2 - 2 R\cos(2\phi) = 1 + R^2 - 2 R(1-2\sin^2(\phi)) = (1-R)^2 + 4R\sin^2(\phi), 
\end{equation}
so that 
\begin{equation}
    T(\phi) = \frac{1}{1 + F \sin^2(\phi)},
\end{equation}
where the coefficient of finesse is 
\begin{equation}
    F = \frac{4R}{(1-R)^2}.
\end{equation}
The transmission probability $T$ is a composition of $f(\phi) = \sqrt{F}\sin\phi$ with the Lorentzian function $L(x) = 1/(1+x^2)$. Several examples of this transmission probability are shown in Fig. \ref{fig:fpi} (a). The finesse parameter is strictly increasing with $R$ in the range $(0, 1)$, and ultimately determines the sharpness of the resonance. The larger the reflection $R$, the more round-trips can occur, on average, before the photon transmits out of the plates. This in turn corresponds to higher phase sensitivity. 

Using the fact that $\phi = 2\pi n d/\lambda$, allows the Fabry-Pérot interferometer wavelength response to be investigated. Plots of $T$ vs. $\lambda$ are shown in Fig. \ref{fig:fpi} (b) for the case $d = 10^{-4}$ meters. The mirror reflection $R$ is assumed to be constant with respect to wavelength. The narrowband response around certain wavelengths makes this device an excellent tool for filtering out a specific wavelength or frequency. Perhaps the most intriguing  physical aspect about this configuration is that the act of aligning two highly reflective plates perpendicular to one another enables certain colors to pass through with \textit{perfect transmission}. This is a result of the interference altering the effective transmission and reflection coefficients of the combined system.

\subsection{Novel higher-dimensional interferometric sensors}

The conventional two-port interferometric configurations shown in Fig. \ref{fig:traditional} employ low-dimensional linear scatterers, namely beam-splitters and plate splitters, in simple graphical configurations. The Sagnac and Michelson devices superimposed two beams, producing a sinusoidal interference pattern. This interferogram was ultimately only sensitive to one parameter in either case, despite the general presence of two. In the Michelson this was the phase difference $\phi_d = \phi_1 - \phi_2$ and for the Sagnac this was the non-reciprocal portion of the loop $\delta$, which is also a relative phase. Meanwhile the Fabry-Pérot configuration, sums an infinite number of paths in a geometric series, resulting in resonances whose width was a function of the plate reflectivities, which are generally fixed. 

\begin{figure}[ht]
    \centering
    \includegraphics[width=0.7\linewidth]{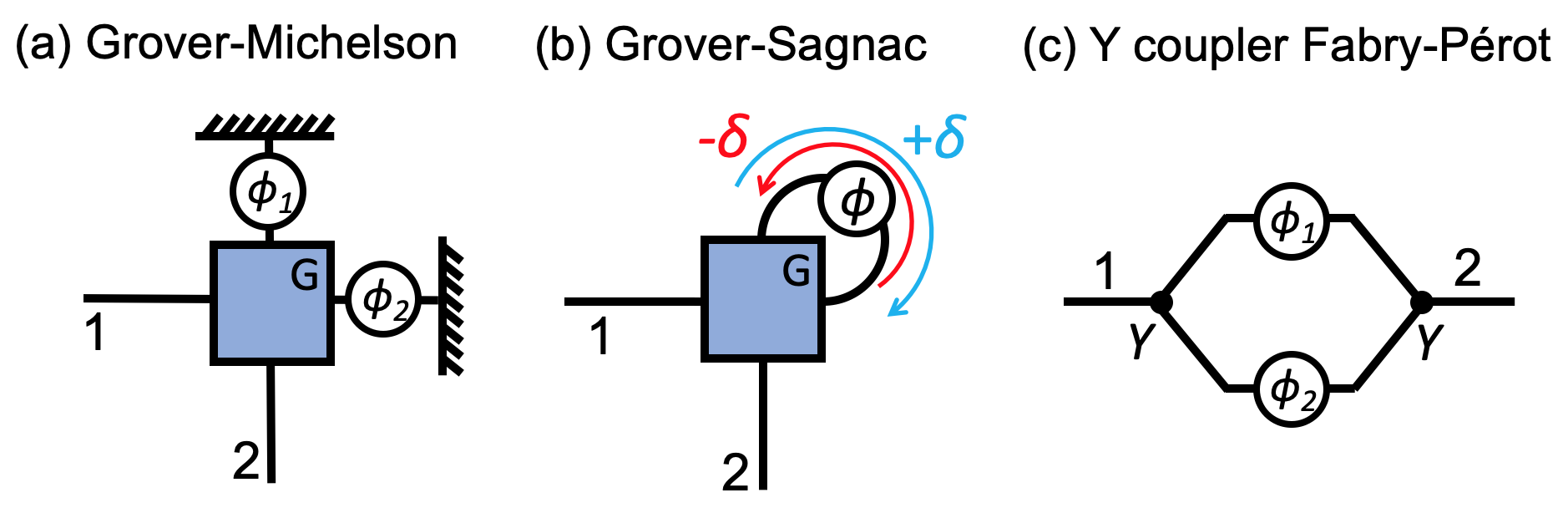}
    \caption{Higher-dimensional analogues of the systems in Fig. \ref{fig:traditional}. The Grover-Michelson (a) and Grover-Sagnac (b) replace the traditional beam splitter with the Grover four-port. In the configuration (c), three-port Y couplers replace the $U(2)$ plate beam splitters of the conventional Fabry-Pérot system. The Grover coin, being fully permutation invariant, does not need to have a specified orientation, but the Y coupler's feed-forward port must be specified. Here we assume that port is the external, unconnected one for each side. In all three of these systems, a pair of coupled optical resonators is formed from the back-reflecting response of the constituent scattering devices. This increase in dimensionality provides access to new degrees of freedom in the system output.}
    \label{fig:higherdim}
\end{figure}

In this section, we introduce a basic generalization to each of these interferometric configurations. We shall see how the generalization to higher dimensions results in each system gaining a parametric degree of freedom. For each case, the constituent scattering devices are replaced with a gauge-symmetric, higher-dimensional equivalent. These configurations are pictured in Fig. \ref{fig:higherdim}. The Grover-Michelson and Grover-Sagnac are formed by replacing the central beam-splitter of their conventional configuration with the four-port Grover coin, while a higher-dimensional Fabry-Pérot interferometer is formed by replacing the two-port plate splitter with the three-port, symmetric Y coupler. The replaced scatterers are directionally unbiased, allowing light to back-scatter. As a result, the single-pass arms of the Michelson are converted to a pair of optical resonators. These cavities are coupled to one other by the Grover coin. The Grover-Sagnac configuration is essentially the same; the two resonators share a common-path inside the loop, corresponding to the two counter-propagating modes of the loop. Because it is a common path configuration, phase differences are only acquired through a non-reciprocal phase $\delta$. The Y-coupler system similarly converts the single-cavity Fabry-Pérot into a double coupled-cavity version.

All of these graphical configurations are highly symmetric. The cavity cross-coupling and out-coupling are equal in each case. Beyond that, each device possesses reflection symmetry with respect to exchange of its external ports. The Y coupler device also has reflection symmetry about the line through those two ports, implying the response is unchanged when $\phi_1$ and $\phi_2$ are exchanged. The same parametric exchange symmetry is found in the Grover-Michelson with $\phi_1$ and $\phi_2$, as well as the Grover-Sagnac, which occurs when the sign of the non-reciprocal phase port $\delta$ is flipped.

\subsubsection{Experimental demonstration of the Grover-Michelson interferometer}

The Grover-Michelson was originally considered in Ref. \cite{PhysRevA.107.052615}. In that work, the output amplitudes were found by studying the normal-mode evolution of the superposition states $a_{c1}^\dagger \pm a_{c2}^\dagger$, where $a_{cj}^\dagger$ is the creation operator for a photon propagating within arm $j$. Defining
\begin{align}
    B &\coloneqq \frac12 (e^{i\phi_1} + e^{i\phi_2}), \\
    C &\coloneqq \frac12 (e^{i\phi_1} - e^{i\phi_2}),
\end{align}
which are precisely the traditional Michelson scattering amplitudes from Eq. (\ref{eq:MI}), the reflection $r$ and transmission $t$ scattering coefficients of the device were found to be
\begin{align}
r &= \bigg (\frac{C^2}{2B - 2} - \frac{B}{2} - \frac12\bigg ),\\
t &= \bigg (\frac{C^2}{2B - 2} - \frac{B}{2} + \frac12\bigg ).
\end{align}
This can be rewritten as
\begin{align}
r &= \frac12 (e^{i\gamma_{\text{GMI}}} - 1) = e^{i\gamma_{\text{GMI}}/2}i\sin(\gamma_{\text{GMI}}/2) \\
t &= \frac12 (e^{i\gamma_{\text{GMI}}} + 1) =e^{i\gamma_{\text{GMI}}/2}\cos(\gamma_{\text{GMI}}/2)
\end{align}
where
\begin{equation}
    \gamma_{\text{GMI}}(\phi_1, \phi_2) = \arctan \bigg ( \frac{\sin(\phi_1 + \phi_2) - \sin\phi_1 - \sin\phi_2 }{\cos(\phi_1 + \phi_2) - \cos\phi_1 - \cos\phi_2 +(1 + \cos(\phi_1 - \phi_2))/2)}\bigg ). 
\end{equation}
\textit{Thus, replacing the beam splitter with a Grover coin can be viewed as a phase transformation of the linear response $(\phi_1 - \phi_2)$ to a synthetic nonlinear one $\gamma_{\text{GMI}}(\phi_1, \phi_2)$.} This is essentially a reparametrization of the general interferometric graph's phase response. 

\begin{figure}[ht]
    \centering
    \includegraphics[width=0.6\linewidth]{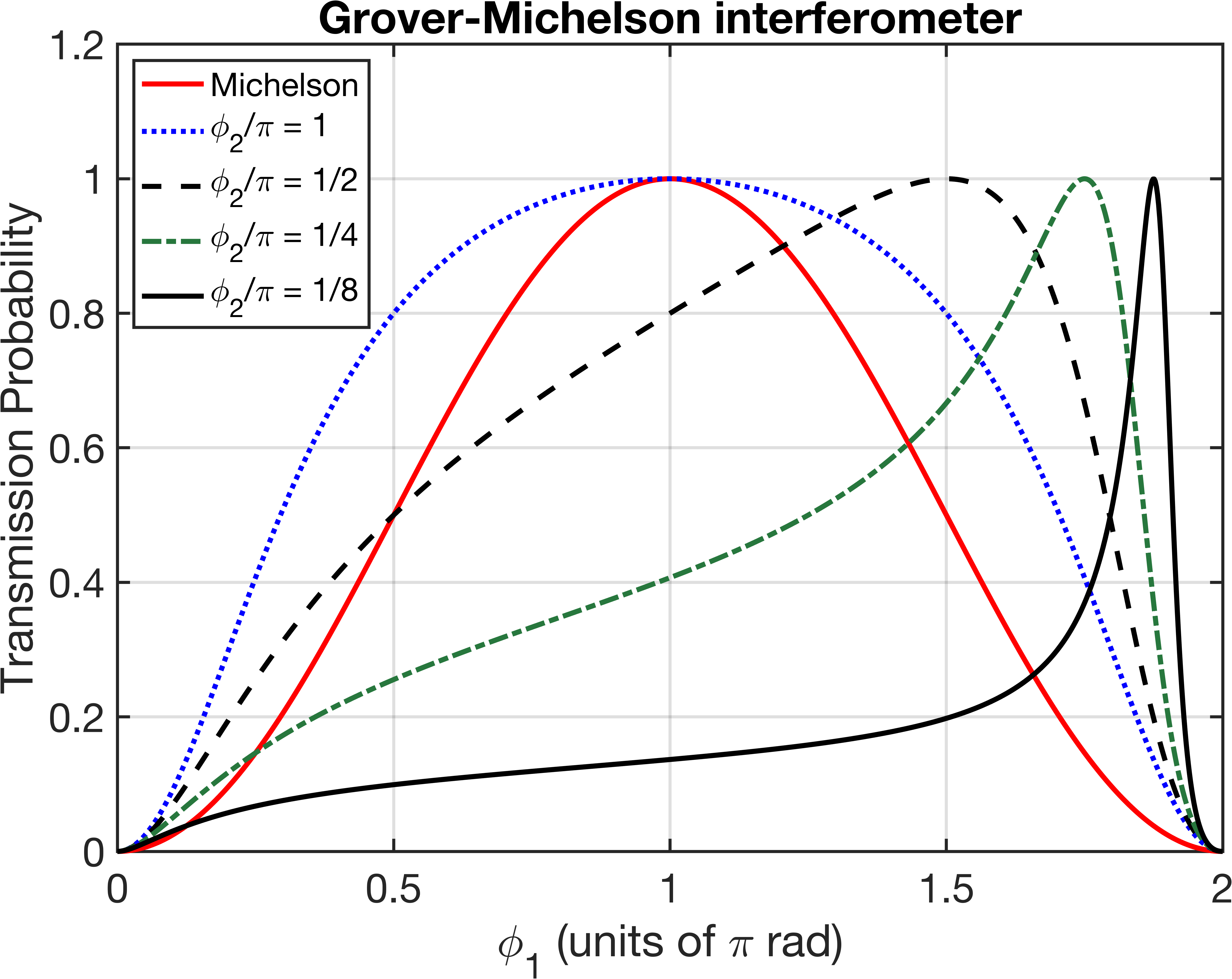}
    \caption{Transmission probability for the Grover-Michelson interferometer. The device is now functions of two parameters, so that the transmission curve generated by sweeping one parameter can be continuously deformed by changing the value of the other parameter. The response can be tuned to exhibit a substantially large slope, which heightens the phase sensitivity. Figure adopted from Ref. \cite{PhysRevA.107.052615}.}
    \label{fig:gmi}
\end{figure}

Example plots of the Grover-Michelson transmission probability are shown in Fig. \ref{fig:gmi}. The probability is plotted with respect to $\phi_1$ for different fixed values of $\phi_2$. The red curve is the traditional Michelson response, for $\phi_2 = 0$. Changing $\phi_2$ merely translates this curve. However, in the Grover-Michelson interferometer, sweeping $\phi_2$ continuously deforms the curves. Being the normal sinusoid but composed with the synthetic phase, the deformation does not affect the visibility, given by $\mathcal{V} = (\max_{\phi_1} T - \min_{\phi_1} T)/(\max_{\phi_1} T + \min_{\phi_1} T)$, which remains fixed at the value of 1 for any $\phi_2$. The position of the transmission peak in $\phi_1$ is given by $2\pi - \phi_2$. As this peak moves towards the value 0 mod $2\pi$, the peak narrows, leading to a heightened phase response in this region. The slope $\partial T/\partial \phi_1$ grows much larger than the traditional Michelson, suggesting a much higher phase sensitivity at the appropriate bias point. 

The decompositions of the Grover coin into two Y couplers (Fig. \ref{fig:decomposition2}) and the Y coupler into a beam splitter and mirror (Fig. \ref{fig:decomposition1}) enable experimental implementation of the Grover coin with a relatively small number of common tabletop optics. As a proof-of-principle, a Grover coin was experimentally assembled like this in bulk and used to form a Grover-Michelson interferometer \cite{Schwarze:24}. The arm phases $\phi_1$ and $\phi_2$ were controlled by modifying their corresponding optical path lengths, which was accomplished by a piezoelectrically actuated translation stage. The source was a quasi-monochromatic, continuous-wave diode laser with an operating wavelength of 633 nm and coherence length exceeding a kilometer. The input power was in the microwatt range; there was no need to reduce the power to the single-photon level since the output obtained is directly proportional to the output probability in that case. Nevertheless, the genuine single-photon implementation would have presented some differences. As the output power is reduced and approaches the single photon level, rather than observing a never-ending continous decrease in registered power as the classical wave picture predicts, the quantized nature of the individual photon energies will eventually manifest as a discrete sequence of clicks, corresponding to the detection of single photons. More experimental details of the setup can be found in Ref. \cite{Schwarze:24}.

\begin{figure}[ht]
    \centering
    \includegraphics[width=0.7\linewidth]{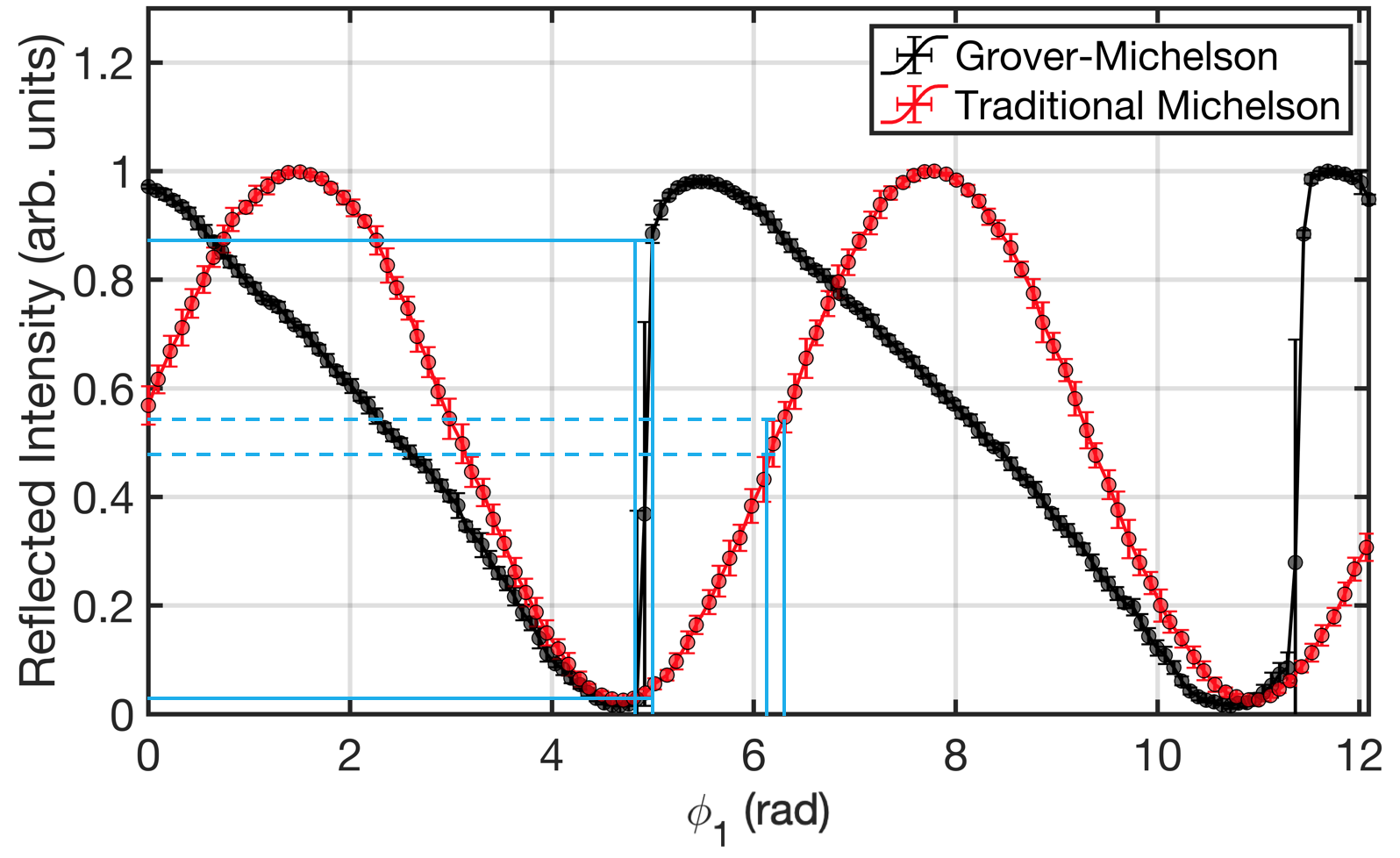}
    \caption{Measured interferogram comparison of a traditional Michelson interferometer (red) and a Grover-Michelson (black). The blue bars loosely depict the difference in intensity modulation for the same input phase modulation of about 145 mrad. The Grover-Michelson exhibited roughly 14X higher intensity change, with a visibility of 97\%. Figure adopted from Ref. \cite{Schwarze:24}.}
    \label{fig:gmie}
\end{figure}

The assembled Grover-Michelson system exhibited curve tunability with high visibility. One example of this, measured from the reflection signal, is overlaid with the measured response of a reference Michelson interferometer. This comparison is shown in Fig. \ref{fig:gmie}. The phase response for the same input phase was notably improved, as loosely depicted by the blue lines. In that case, where $\Delta\phi_1 \approx 145$ mrad,\textit{ the corresponding intensity modulation is about 14 times larger for the Grover-Michelson interferometer than its traditional counterpart. } This might further improved by reducing losses. Other modifications to the system could also be considered, as in Ref. \cite{Wei:25}

\subsubsection{Grover-Sagnac interferometer}

\begin{figure}[ht]
    \centering
    \includegraphics[width=0.6\linewidth]{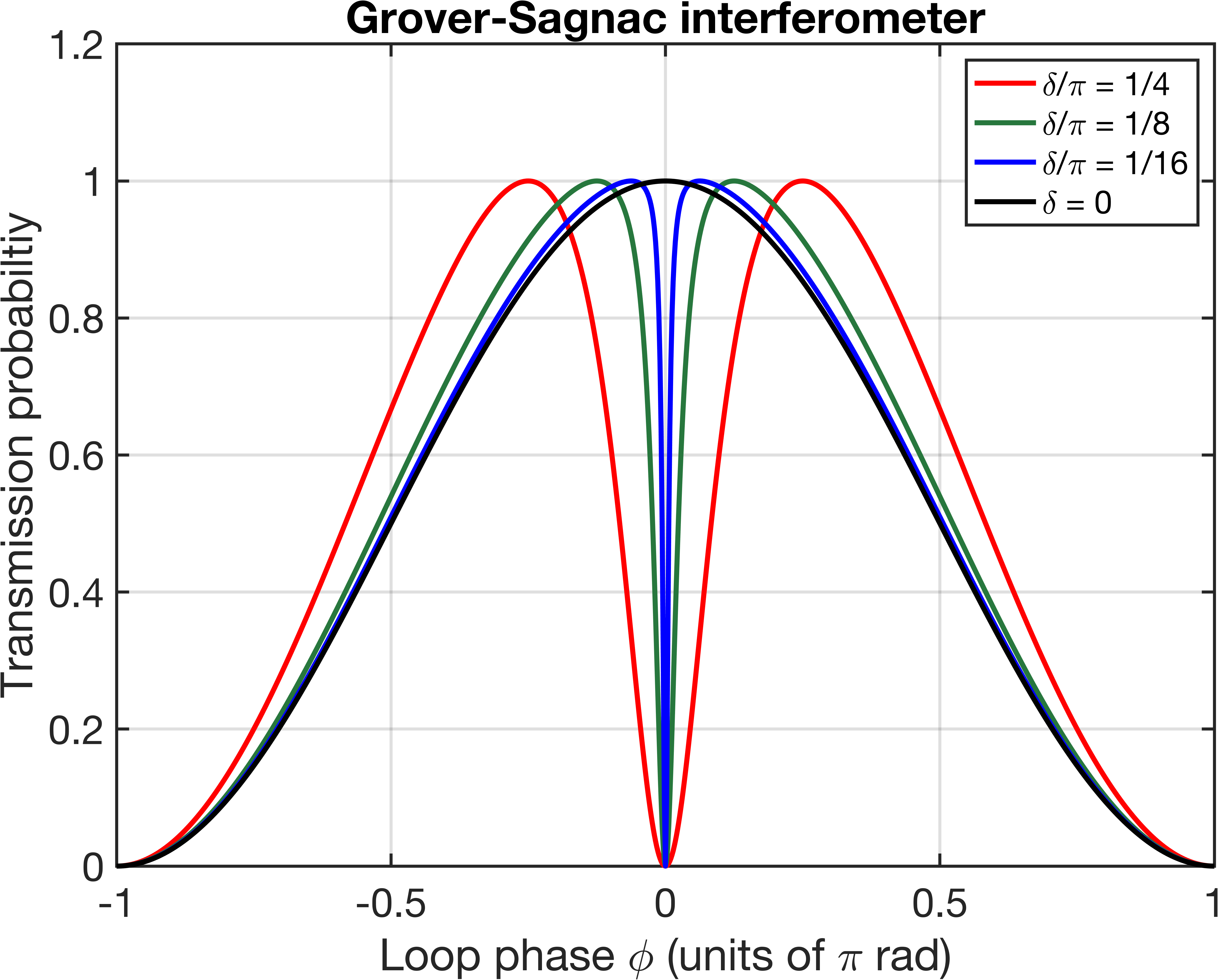}
    \caption{Transmission probability of the Grover-Sagnac interferometer. Unlike in the traditional Sagnac interferometer, the probability is a continuous function of both $\phi$ and $\delta$. For $\delta = 0$ the curve is sinusoidal in $\phi$, but exhibits resonance whenever $\delta \neq 0$. This leads to the formation of a sharp dip about the origin $\phi = 0$. The full width of this dip is $2\delta$, enabling a high-contrast, direct readout of $\delta$. Figure adopted from Ref. \cite{Schwarze:25}.}
    \label{fig:gsi}
\end{figure}

A similar analysis has been be applied to the Sagnac configuration \cite{Schwarze:25}. Instead of instead of $\gamma_{\text{GMI}}$, the phase function
\begin{equation}
    \gamma_{\text{GSI}}(\phi, \delta) = \phi - \arctan \bigg (\frac{\sin\phi\sin^2\delta}{2\cos\delta - \cos\phi (1 + \cos^2\delta)} \bigg )
\end{equation}
is obtained. Plots of the associated transmission are shown in Fig. \ref{fig:gsi}. When $\delta = 0$, $\gamma_{\text{GSI}} = \phi$. Thus the reflection and transmission probabilities depend on the reciprocal part of the loop phase, $\phi$, and can take on any output value. This is not the case in a traditional Sagnac interferometer, where the lack of non-reciprocal phase forces the transmission to be zero regardless of the reciprocal loop phase. \textit{Moreover, the resonance, or pole in the phase response, occurs around the point $\delta = 0$. This means small values of $\delta$ create a sharp dip in the transmission of full-width $2\delta$.} Scanning $\phi$ around the origin provides a direct readout for this in the interferogram. This signature is linear in $\delta$ for any $\delta$, whereas the traditional Sagnac interferometer's transmission is given by $\delta^2$ for small $\delta$, placing constraints on the detector resolution. The full-modulation dip is easy to detect, but requires inducing a change in phase $\phi$ which is significantly lesser than $\delta$ in order to resolve $\delta$ with high precision. 

\subsubsection{Higher-dimensional Fabry-Pérot interferometer}

The output of the Y-coupler Fabry-Pérot device pictured in Fig. \ref{fig:higherdim} also can be expressed in terms of that of a regular Michelson interferometer \cite{PhysRevA.110.023527}. The output is given by
\begin{subequations}
\begin{align}
    r &= -\frac{C^2}{1 - B^2},\\
    t &= B(1-r).
\end{align}
\end{subequations}

\begin{figure}[ht]
    \centering
    \includegraphics[width=0.6\linewidth]{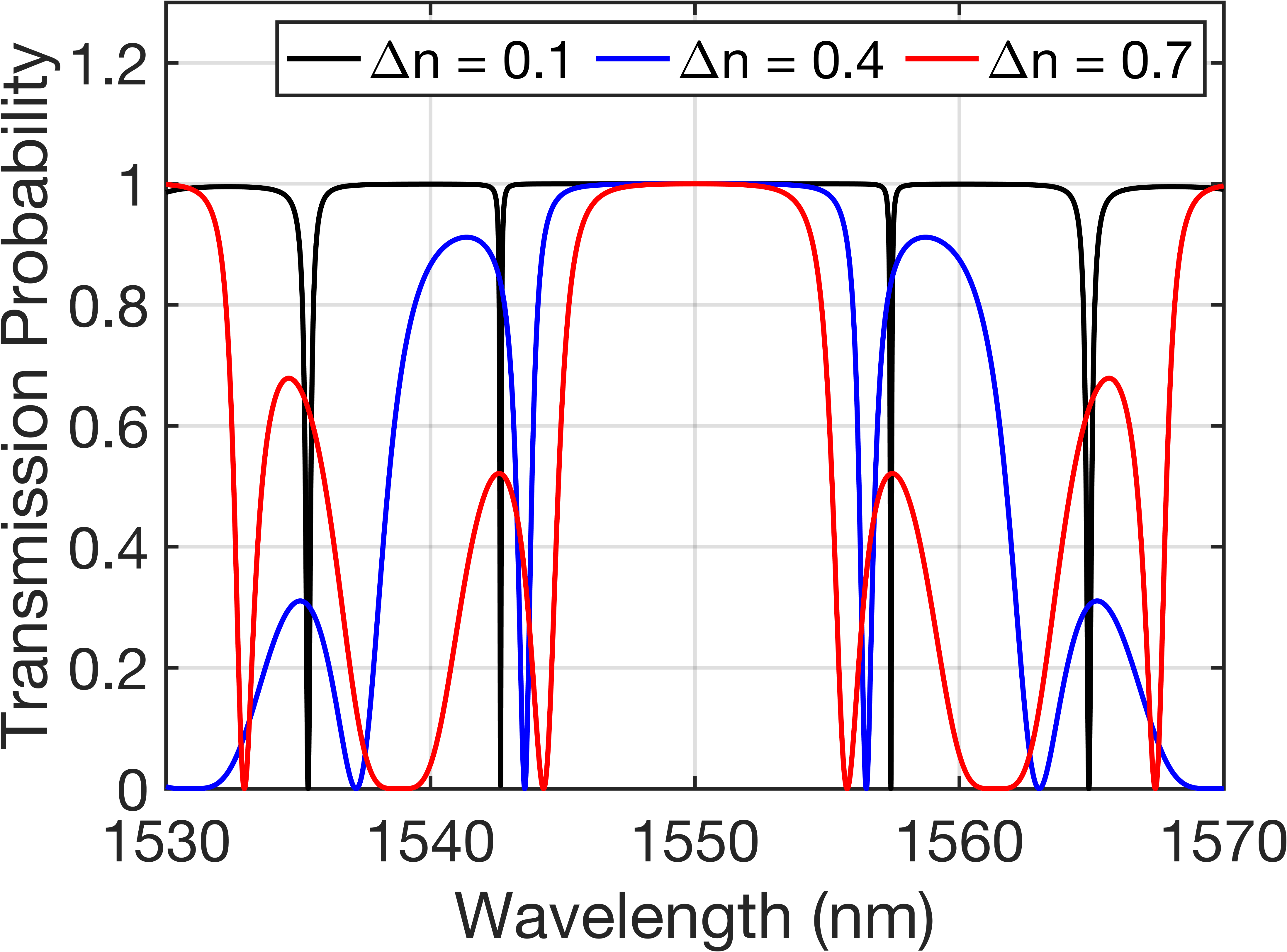}
    \caption{Example output curves for the Y coupler Fabry-Pérot interferometer. The phase was modulated by perturbing the refractive index in one arm relative to the other. The arms were each assumed to be 155 um long. Changing the relative phase acquired by each arm can bring certain bands from high transmission to high reflection, acting as a tunable dichroic. } 
    \label{fig:ycmzifpi}
\end{figure}

The traditional Fabry-Pérot only has one phase parameter, which for symmetric plates, traced a Lorentzian-like function whose sharpness was determined by the finesse $F$. $F$ is a function of the plate reflectivity, which is typically a quantity fixed by material properties. \textit{In the higher-dimensional Y-coupler version, there are now two phases. The additional degree of freedom allows tuning of the effective out-coupling between the cavity and the output ports.} Thus the effective finesse is tunable through interference. This allows the widths and heights of certain spectral regions to be controlled with considerable flexibility. A basic example is shown in Fig. \ref{fig:ycmzifpi}. The arms are each assumed to be $1.55\cdot10^{-4}$ meters with a refractive index of 1. By perturbing the index of one relative to the other, a phase difference is created between the two arms. Increasing this index perturbation completely reconfigures the filter characteristics, allowing transmission bands to be converted to reflection bands. Overall, the device acts as a tunable dichroic. 

\subsection{Hong-Ou-Mandel two-photon interference}

\begin{figure}[ht]
    \centering
    \includegraphics[width=\linewidth]{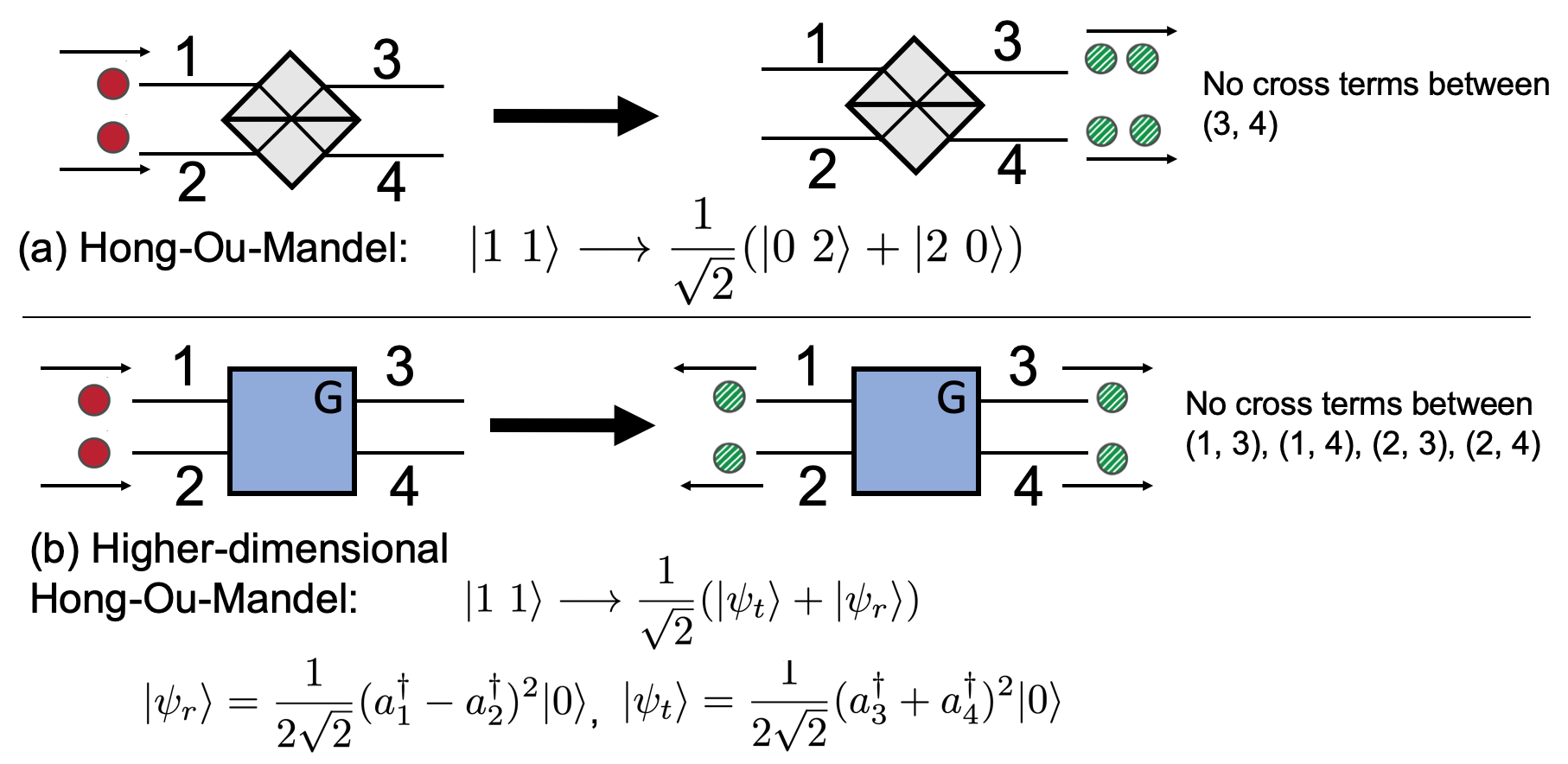}
    \caption{(a) The traditional Hong-Ou-Mandel effect. When a pair of identical photons enter different ports of a balanced unitary beam splitter, the coincident amplitude corresponding to one photon populating each of the two output modes destructively interferes. The output state is then comprised of two clusters of two photons each. Each cluster occupies a single mode. (b) Higher-dimensional Hong-Ou-Mandel effect. When the two photons impinge a Grover coin, clusters form again, with destructive interference occurring across four pairs of output modes. The resulting clusters are each distributed over two modes instead of the usual one. This increase in the dimensionality enables the energy in each cluster to be routed into one of two different output modes.}
    \label{fig:hom}
\end{figure}

Up to this point, all of the input states have been single-photon states, which only exhibit classical interference. In this section we consider two-photon states entering separate ports of an optical scattering device, which are of the form $|\psi_{\text{in}}\rangle = a_1^\dagger a_2^\dagger |0\rangle$. The resulting two-photon interference can be measured with multiple detectors placed at the output ports which count the photons that arrive simultaneously. Perfect destructive interference of the underlying coincident amplitude leads to a famous Hong-Ou-Mandel (HOM) dip in the rate of coincidences \cite{HOM, Drago_2024}. The situation is shown in Fig. \ref{fig:hom} (a). Two identical photons impinge separate ports of a generic unitary beam splitter of the form in Eq. (\ref{eq:BS}). Each individual photon sees its normal scattering action, resulting in the transformation
\begin{equation}
    a_1^\dagger a_2^\dagger|0\rangle \rightarrow (r_1 a_3^\dagger + t_1 a_4^\dagger)(r_2 a_4^\dagger + t_2 a_3^\dagger)|0\rangle. 
\end{equation}
The coincident amplitude, which leads to the $a_3^\dagger a_4^\dagger$ term, is then $r_1r_2 + t_1t_2$. From the structure of the beam splitter, we recall the amplitudes $r_1, r_2, t_1$ and  $t_2$ originate from a two-dimensional unitary transformation, which can be written in the form of Eq. (\ref{eq:U2}), in which the coefficients are rewritten as separated magnitudes and phases. 
A consequence of unitarity in two dimensions was the phase constraint of Eq. (\ref{eq:HOM}), which states that the difference between the summed phases of the $r_j$ and the summed phases of the $t_j$ had to be $\pi$. This is precisely the relative phase between the two terms comprising the coincident term. After dropping the irrelevant global phase, then, this coincident term can be rewritten as $|r_1||r_2| - |t_1||t_2|$. Accordingly, if and only if the unitary beam splitter has a 50:50 splitting ratio, the coincident term vanishes. This is the origin of the HOM effect, producing an output state $(a_3^{2\dagger} + a_4^{2\dagger})|0\rangle$. The produced state is of two \textit{clustered} photon amplitudes. The clusters each populate a single mode. There is an equal probability of measuring a cluster in either of the two modes. 

\textit{A higher-dimensional generalization to this occurs when two photons impinge two ports of a Grover coin \cite{PhysRevA.102.063712}.} This situation is depicted in Fig. \ref{fig:hom} (b). The state produced is 
\begin{equation}
    a_1^\dagger a_2^\dagger \rightarrow \frac14 (-a_1^\dagger + a_2^\dagger + a_3^\dagger + a_4^\dagger)(a_1^\dagger - a_2^\dagger + a_3^\dagger + a_4^\dagger)|0\rangle.
\end{equation}
Another cancellation of cross terms occurs, since $(-a_1^\dagger + a_2^\dagger + a_3^\dagger + a_4^\dagger)(a_1^\dagger - a_2^\dagger + a_3^\dagger + a_4^\dagger) = (a_1^\dagger - a_2^\dagger)^2 + (a_3^\dagger + a_4^\dagger)^2 + (a_3^\dagger + a_4^\dagger)(a_1^\dagger - a_2^\dagger) - (a_3^\dagger + a_4^\dagger)(a_1^\dagger - a_2^\dagger)$. The final state is also clustered into two components, 
\begin{equation}\label{eq:HDHOM}
    |\psi\rangle =  \frac{1}{\sqrt{2}}(|\psi_r\rangle + |\psi_t\rangle). 
\end{equation}
The reflected component is $|\psi_r\rangle =(1/2\sqrt{2})(a_1^\dagger - a_2^\dagger)^2|0\rangle$ whereas the transmitted component is $|\psi_t\rangle =(1/2\sqrt{2})(a_3^\dagger + a_4^\dagger)^2|0\rangle$. These symmetric state components are two-photon eigenstates of the Grover coin: $G|\psi_t\rangle = |\psi_t\rangle$, while $G|\psi_r\rangle = -|\psi_r\rangle$.

The output state (\ref{eq:HDHOM}) encapsulates \textit{simultaneous destructive interference of coincident amplitudes over four mode pairs}: $(1, 3), (1, 4), (2, 3)$, and $(2, 4)$. These are the pairs on opposite sides of the Grover coin shown in Fig. \ref{fig:hom}. The clustered components are also now distributed over two modes instead of the usual one. This readily enables routing the energy spread across each cluster to be concentrated into a single mode. There are two routing options per each of the two clusters, allowing the emergent two-photon amplitudes to take on the same form as the traditional Hong-Ou-Mandel effect but with four different switchable options. The effect has since been shown to enable the complete routing of entangled states in a general quantum network \cite{PhysRevA.102.063712, PhysRevA.104.012617}.

If the two photons in either Hong-Ou-Mandel configuration enter the device with a time delay between them, the delay introduces temporal distinguishability which diminishes the perfect destructive interference \cite{Drago_2024, Meany:12}. Varying the delay allows the degree of distinguishability to be tuned, inducing a change in the coincident rate. As the delay passes through the zero point, perfect destructive interference occurs in a sharp dip, which can be used to measure small time delays or the degree of indistinguishability of the source. \textit{Because of the fourfold cancellation of cross-terms in the higher-dimensional Hong-Ou-Mandel configuration, four simultaneous dips will occur.} The relative positions of each dip and peak will be given by the difference in phases acquired after the photons exit the Grover coin. When the phases are equal, the dips will overlap, allowing the simultaneous determination of the three relative phases. These are defined across a given port and the other three relative to that. Sensing multiple phase shifts is a general benefit of multimode interference systems. Another example will be considered in the next section. 

\subsection{Two-photon Grover-Mach-Zehnder interferometry}
\begin{figure}[ht]
    \centering
    \includegraphics[width=\linewidth]{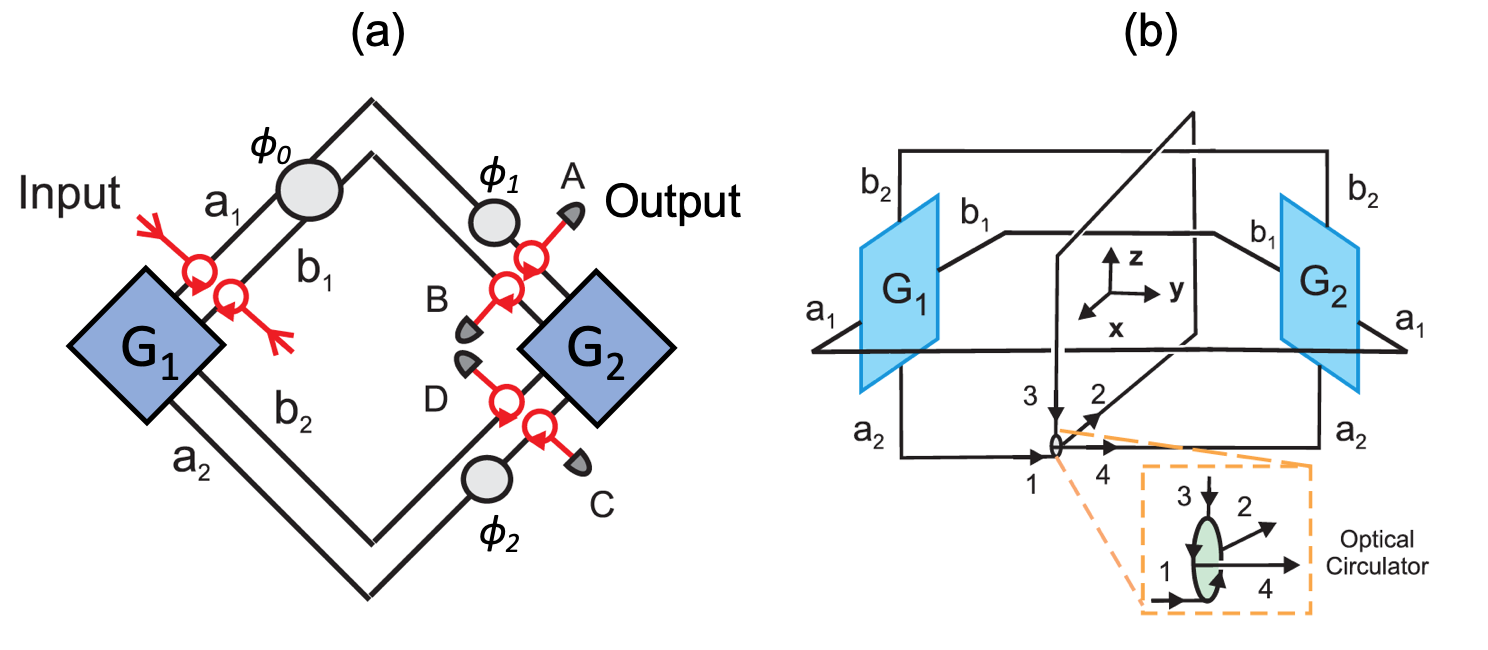}
    \caption{(a) Dual-rail Grover-Mach-Zehnder configuration. Two photons are injected at the input via circulators, scatter off a Grover coin $G_1$ and acquire the phases $\phi_0, \phi_1$ and $\phi_2$ while traversing the two double-rail arms. After this the optical state scatters off another Grover coin $G_2$ and then are out-coupled to detectors $A, B, C, D$ with circulators. By recording the coincidences of each detector pair, the three phases can be determined simultaneously. (b) By introducing a circulator into the dual-rail arms and placing the arms in different planes, the acquired phases can made to depend on the rotation rates of the different axes, as described by Eq. (\ref{eq:rot}). This allows the system to simultaneously detect rotations about three axes. Figure adopted from Ref. \cite{PhysRevA.106.033706}.}
    \label{fig:gmzis}
\end{figure}

A double-rail generalization of the Mach-Zehnder interferometer (shown in Fig. \ref{fig:mimzi}) can be obtained by replacing the two beam-splitters with Grover coins and linking the two unconnected ports to obtain the configuration in Fig. \ref{fig:gmzis} (a). Circulators in red are used to inject two photons into the first Grover coin, $G$. The emergent state is of the form (\ref{eq:HDHOM}). After it emerges from $G_1$, the state acquires the phases $\phi_0, \phi_1$ and $\phi_2$, scatters from a second Grover coin $G_2$, and then exits the system with another set of circulators. The state is detected across four detectors, labeled $A, B, C$ and $D$.\footnote{Since the circulators are only used to separate input from output, the same configuration could be achieved without them if block-feed forward Grover coins were used. This feed-forward Grover matrix is equivalent to the system pictured in Fig. 5 of Ref. \cite{PhysRevA.93.012323}. }

The coincidence rates of each detector pair are readily extracted from the output state amplitudes \cite{PhysRevA.106.033706}. For $i, j$ in $\{A, B, C, D\}$, denote the coincidence rate between detectors $i$ and $j$ as $R_{ij}$. These rates can be related to the phases $\phi_0, \phi_1$ and $\phi_2$, 
\begin{subequations}
\begin{align}
    \frac{R_{AC} + R_{AD}}{2R_0} &= \sin^2\phi_1 + \sin^2\phi_2\\
    \frac{R_{AD} + R_{AC}}{2R_0} &= 2\sin\phi_1\sin\phi_2\cos(2\phi_0+\phi_1 - \phi_2)\\
    \frac{R_{AB} + R_{CD}}{2R_0} &= (1+\cos^2\phi_1) + (1+\cos^2\phi_2) + 2\cos(2\phi_0 + \phi_1 - \phi_2)(1+\cos\phi_1\cos\phi_2)\\
    \frac{R_{AB} + R_{CD}}{2R_0} &= 2(\cos\phi_1+\cos\phi_2)(1+\cos(2\phi_0 + \phi_1 - \phi_2)).
\end{align}
\end{subequations}
This set of four independent equations allows the three phases and the constant $R_0$ to be obtained simultaneously. In the case of two independent phase shifts, i.e. $\phi_1 = \phi_2$ the inversion of the above equations becomes direct, for in this case,  
\begin{subequations}
\begin{align}
    \cos2\phi_0 &= \Lambda_1\\
    \cos\phi_1 &= \frac{1 \pm \sqrt{1 - \Lambda_1^2}}{\Lambda_2}
\end{align}
\end{subequations}
where
\begin{subequations}
\begin{align}
\Lambda_1 &= \frac{R_{AD} - R_{AC}}{R_{AD} + R_{AC}}\\
\Lambda_2 &= \frac{R_{AB} - R_{CD}}{R_{AB} + R_{CD}}.
\end{align}
\end{subequations}

The parameters $\phi_0, \phi_1$ and $\phi_2$ are general phase shifts. If the system were rotated, these phases could be shifted in conjunction with the Sagnac effect. Placing the arms in different planes allow rotations about different axes to affect the different phase shift parameters. By adding a circulator to the loops as shown in Fig. \ref{fig:gmzis} (b), a third loop can be obtained, which allows all three phases to be tied to a rotation direction. With the circulator added, the path from $G_1$ to $G_2$ exiting $a_1$ or $b_1$ comprises a half-loop about the $z$ axis, the path leaving $b_2$ comprises a half-loop about the $x$ axis, and each path leaving $a_2$ comprises both a full loop about $y$ and a half loop about $x$. This allows the phase shifts $\phi_0, \phi_1$, and $\phi_2$ above to be expressed in terms of the phase shifts induced by rotations about the $x, y$ and $z$ axes like so:
\begin{subequations}\label{eq:rot}
\begin{align}
    \phi_0 &= (\phi_x - \phi_z)/2,\\
    \phi_1 &= \phi_z, \\
    \phi_2 &= \phi_x + \phi_y.
\end{align}
\end{subequations}
Given measurement of $\phi_0, \phi_1$ and $\phi_2$, the above equations are readily inverted to obtain the rotation phases, 
\begin{subequations}
\begin{align}
    \phi_x &= 2\phi_0 + \phi_1\\
    \phi_y &= -2\phi_0 - (\phi_1 + \phi_2)\\
    \phi_z &= \phi_1.
\end{align}
\end{subequations}
Thus the system offers the simultaneous determination of three rotations about different axes.

\section{Discussion}
In this review, we have discussed the design principles of novel gauge-invariant optical devices capable of providing a substantial boost in resolution of interferometric sensors. A general approach for identifying physically-equivalent scatterers has been presented, which provides a basis to recognize or enforce certain symmetries in a scattering transformation. An emphasis was placed on the gauge-invariant, higher-dimensional Y-coupler and Grover coins, due to their role in interferometry. The general matrix theory for optical scattering has several other extensions and applications, such as unitary control \cite{PhysRevA.111.023507, PhysRevB.110.035430, PhysRevB.110.035431}. Certain assumed global symmetries can also be used to bound the elements of a scattering matrix \cite{8bcn-796c, PhysRevLett.128.256101}. 

Although the Grover coin and Y-coupler were constructed through a beam-splitter decomposition, there has yet to be a demonstration of a passive, single-element implementation of these devices, in any platform. Such an implementation would be valuable to reduce the number of resources used in larger systems. A variety of beam-splitters have been realized using metasurface technology, making this a potential candidate for realizing higher-dimensional unbiased multiports in the future \cite{Wu2024, Hemayat:23, nanomanufacturing2040014, ms1, ms2}. Other possible approaches include integrated multi-mode interference (MMI) devices \cite{372474, 740707, 300172, Maese-Novo:13, Bachmann:94, Cooney:16} and nanophotonic inverse design \cite{molesky2018, KangParkLeeKangJangChung, huang2025multiphotonquantuminterferenceultracompact}. These approaches all involve reducing the scale of the optical device. This would be advantageous in situations where the coherence length of the source was no longer large enough to be ignored to a good approximation. When the coherence length becomes finite, as in the case of a pulsed or broadband white light source, the interference of field amplitudes will no longer occur when the difference in optical path lengths becomes greater than the coherence length. Even if the individual path components are short, devices involving multiple bounces with low loss can obtain a large path length in effect. Here this potential problem was avoided with the perfect coherence which is furnished by an assumed monochromatic source. Nevertheless, this configuration is only employed in some optical interferometric systems, so it remains to be seen how other sources with lower coherence may be employed for unbiased multiport interferometry, and the effect of difference system scales on the output.

All systems reviewed were also assumed to be unitary. However, non-unitary systems with loss are a practical reality stemming from artifacts such as absorption or undesired radiating out of the system under consideration. Although a non-unitary system can always be cast into a higher dimensional unitary space to account for inaccessible "lossy modes," the non-unitary systems themselves have become increasingly popular area of research. This is because in general a non-unitary system is not bound to the same limits as a unitary scatterer, which enables other new effects to be observed \cite{s22113977, PhysRevResearch.2.013280, YanZhaoZhouMaLyuChuHuGong, PartoLiuBahariKhajavikhanChristodoulides, Longhi2015, Li2023}.

Interferometers were defined as graphs of static scattering devices, producing a scatterer which is a function of tunable parameters such as induced phase shifts. After discussing traditional interferometric systems, higher-dimensional systems built around the Grover coin and Y coupler were presented. These scatterers utilize more field modes per scattering event, exhibiting rich interference effects which directly translated to practical improvements for interferometric phase sensors. These improvements included higher resolution, more control freedoms, and the ability to measure several phases simultaneously. Naturally, the systems employing coupled resonances such as the Grover-Michelson interferometer will tend to exhibit a smaller range where the slope is increased. This local narrowing of the response range requires an active phase controller to bias the system at the sensitive point. Other applications which may benefit from the increased sensitivity exhibited in these systems include optical switches and modulators, phase contrast microscopes, optical filters.

The enhancements resulted directly from the replacement of a lower-dimensional scattering device with a higher-dimensional one in the same graphical configurations. It remains to be seen what effects other configurations and inputted optical states may produce in these symmetric interferometers.

\section*{Author contributions}
Conceptualization, C.R.S., A.D.M., D.S.S., A.N. and A.V.S.; writing—original draft preparation, C.R.S.; writing—review and editing, C.R.S., A.D.M, D.S.S., A.N., A.V.S.; supervision, A.V.S.; funding acquisition, A.V.S.

\section*{Funding}
This research was funded by Air Force Office of Scientific Research MURI award number FA9550-22-1-0312.

\section*{Data availability}
No new data were created or analyzed in this study.

\section*{Conflicts of interest}
The authors declare no conflicts of interest.

\bibliography{your_external_BibTeX_file}

\end{document}